\documentclass{cta-author}
\usepackage{hyperref}
\hypersetup{hypertex=true,
	colorlinks=true,
	linkcolor=blue,
	anchorcolor=blue,
	citecolor=blue}
\usepackage{subfigure}
\usepackage{longtable}
\usepackage{multirow}
\usepackage[normalem]{ulem}
\usepackage{longtable}
\useunder{\uline}{\ul}{}
\usepackage{hhline}
\usepackage{graphicx}
\usepackage{url}

\usepackage{pifont}
\usepackage{array}

{}
{}
{}

\begin{document}

\supertitle{Submission Template for IET Research Journal Papers}

\title{Blockchain for Finance: A Survey}

\author{\au{Hanjie Wu$^{1,2}$}, \au{Qian Yao$^{3\corr}$}, \au{Zhenguang Liu$^{1,2}$}, \au{Butian Huang$^{4}$}, \au{Yuan Zhuang$^{5}$}, \au{Huayun Tang$^{6}$}, \au{Erwu Liu$^{7}$}}

\address{\add{1}{School of Cyber Science and Technology, Zhejiang University, Hangzhou, China}
\add{2}{ZJU-Hangzhou Global Scientific and Technological Innovation Center, Hangzhou, China}
\add{3}{China Securities Regulatory Commission, China}
\add{4}{Hangzhou Yunphant Network Technology Co. Ltd., Hangzhou, China}
\add{5}{College of Computer Science and Technology, Harbin Engineering University, Harbin, China}
\add{6}{Institute of Chinese Finance Studies, Southwestern University of Finance and Economics, Chengdu, China}
\add{7}{College of Electronics and Information Engineering, Tongji University, Shanghai, China}
\email{yaoqian@csrc.gov.cn}}

\begin{abstract}
As an innovative technology for enhancing authenticity, security, and risk management, blockchain is being widely adopted in trade and finance systems. The unique capabilities of blockchain, such as immutability and transparency, enable new business models of distributed data storage, point-to-point transactions, and decentralized autonomous organizations. In this paper, we focus on blockchain-based securities trading, in which blockchain technology plays a vital role in financial services as it ultimately lifts trust and frees the need for third-party verification by using consensus-based verification. 
We investigate the 12 most popular blockchain platforms and elaborate on 6 platforms that are related to finance, seeking to provide a panorama of securities trading practices.
Meanwhile, this survey provides a comprehensive summary of blockchain-based securities trading applications. We gather numerous practical applications of blockchain-based securities trading and categorize them into four distinct categories. For each category, we introduce a typical example and explain how blockchain contributes to solving the key problems faced by FinTech companies and researchers. 
Finally, we provide interesting observations ranging from mainstream blockchain-based financial institutions to security issues of decentralized finance applications, aiming to picture the current blockchain ecosystem in finance. 

\end{abstract}

\maketitle

 \section{Introduction}\label{sec1}

Fueled by the rapid increase of information technology in the past decades, blockchain, \textit{considered one of the most impactful innovations,} was proposed by Nakamoto in 2008 \cite{ref10}. A blockchain is essentially a distributed transaction ledger, shared by miners in the blockchain system following consensus protocols \cite{ref127}. The replicated ledger shared by numerous miners and consensus protocol enforce all transactions immutable once written in the ledger, endowing blockchain with immutability and decentralization nature \cite{ref127}. 

Blockchain technology has paved the way for novel applications in \textit{data sharing, data security,} and \textit{trading}. Blockchain technology has captivated significant industry investments, due to its \textit{decentralization, safety,} and \textit{traceability} nature. At present, numerous applications of blockchain across a wide range of sectors are being implemented, including healthcare \cite{ref115}, IoT \cite{ref116}, data privacy \cite{ref117}, supply chain \cite{ref118}, goods tracing \cite{ref119}, energy management \cite{ref120}, and combating product counterfeiting \cite{ref121}, etc. Among the various sectors in blockchain applications, \textit{FinTech} has emerged as a prominent and promising area for exploration \cite{ref122}.

The global financial system provides services to billions of people daily while managing trillions of cash \cite{r1}\cite{r2}\cite{r3}. To illustrate this, Fig.\ref{F} presents a portrayal of the global debt securities market across the preceding seven years. In this expansive market landscape, traditional financial infrastructures rely upon established third-party entities to cultivate and sustain trust. Nonetheless, this prevailing model comes with its inherent challenges. These challenges mainly encompass the expenditure of having numerous stakeholders, the persistent issue of delays, the cumbersome burden of excessive paperwork, and the ever-looming threat of data breaches \cite{r4}\cite{r5}\cite{r6}. The cumulative impact of these challenges translates into \textit{high cost, low efficiency,} and \textit{frequent security issues} \cite{ref7}\cite{ref8}\cite{ref9}.

However, the landscape of financial behaviors, including banking and trading, has transformed since the emergence of blockchain \cite{ref125}.
Blockchain technology has the potential to address the above issues in financial areas. 
This potential emanates from blockchain's distinctive characteristics, namely \textit{decentralization, multi-party bookkeeping,} and \textit{immutability} \cite{r2}\cite{r4}. 
The robustness and efficiency of financial systems can be improved through blockchain's decentralized management strategy, particularly in the context of the securities market \cite{r3}. Using blockchain technology in the securities market, the high costs incurred by intermediaries such as \textit{regulatory agencies, brokers,} and \textit{stock exchanges} can be mitigated. Hence, the attainment of decentralization within the securities market represents a pivotal progression. Central to this shift lies the distributed trust inherently embedded in blockchain technology, which catalyzes the financial revolution from three aspects: (1) eliminating reliance on trusted third parties, (2) decreasing the cost of trading, and (3) reducing the time delay \cite{ref65}\cite{ref66}\cite{r1}\cite{r6}\cite{ref7}.  

In this paper, we focus on blockchain applications in financial areas, especially on applications for the securities market. There are many challenges in traditional securities trading, including inefficient securities trading, low liquidity of financial assets, security issues, and so on. Due to the advantages of blockchain technology, it is considered a promising solution for the above challenges. 

Despite many high-level reviews of blockchain technology being presented, a systematic analysis of blockchain applications within financial areas is still lacking.
Motivated by this, in this survey, we study and summarize the existing literature on blockchain applications, aiming to explore the strengths and weaknesses of blockchain when applied to financial areas.
Specifically, we spare more efforts on securities-related blockchain applications. 
Technically, we categorize the applications into four categories. 
For each category of applications, we introduce a typical example of how blockchain contributes to solving the problems in a specific finance sub-area. 
We observe that most blockchain applications are built upon 12 popular blockchain platforms. We introduce the different underlying implementations of these blockchain platforms in terms of consensus protocols and smart contracts.
In addition, interesting observations ranging from mainstream blockchain-based financial institutions to security issues of decentralized finance applications are also presented, aiming to picture the current blockchain ecosystem in finance.

The remainder of this article is structured as follows: Section 2 gives a brief introduction to the background knowledge of traditional financial infrastructure and blockchain technology. Section 3 presents financial blockchain platforms and analyzes the advantages of implementing financial services with these platforms. Section 4 divides blockchain applications related to securities trading into four categories and elaborates on existing practices for each category. Section 5 lists several interesting observations, including a comparative analysis of current blockchain platforms, trends of mainstream blockchain-based financial institutions, and security issues of decentralized finance applications. Finally, Section 6 concludes and provides suggestions for future research.

 \begin{figure*}
		\includegraphics[scale=0.6]{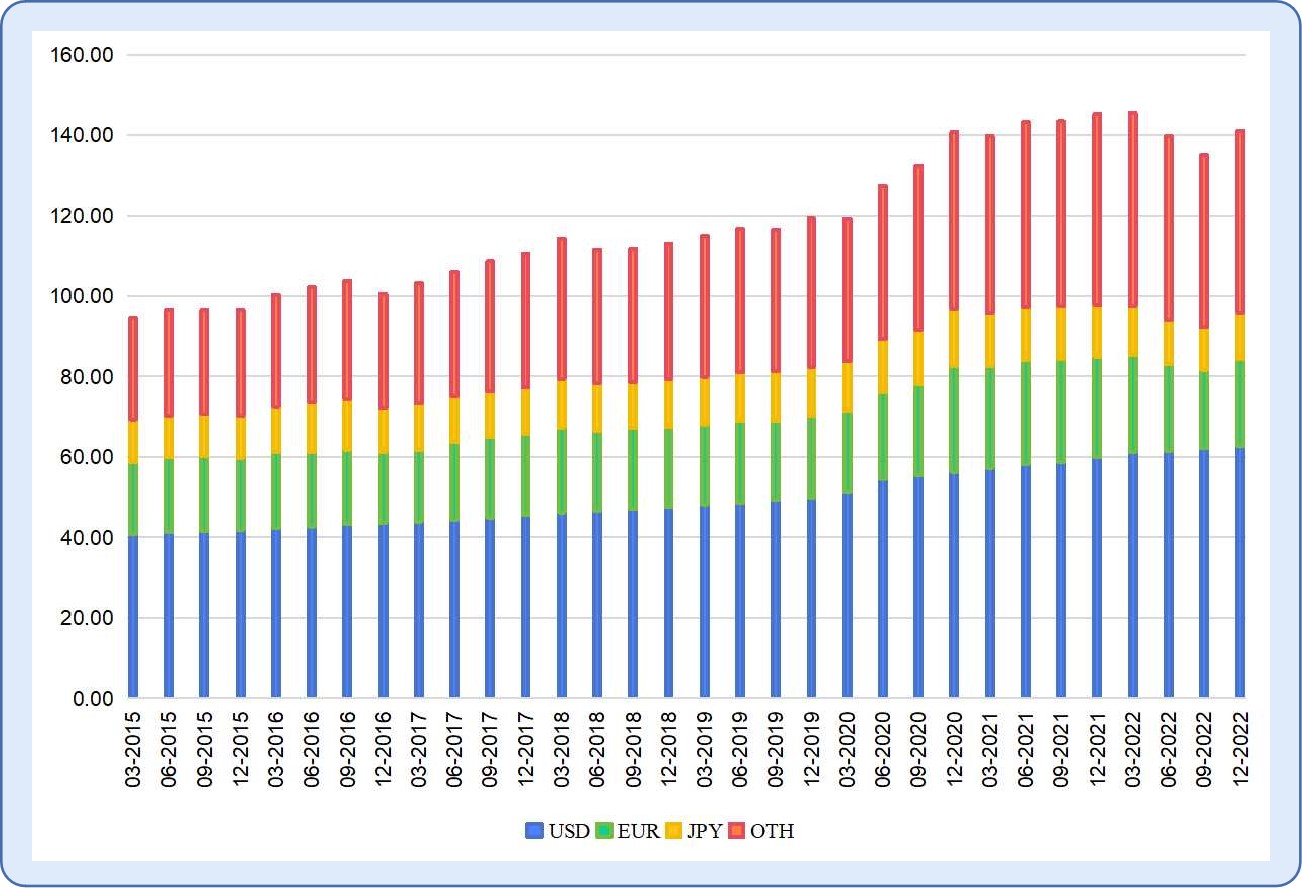}
		\centering
		\caption{\textbf{The size of global debt securities markets}\protect\footnotemark[1]\textbf{(By currency of denomination}\protect\footnotemark[2]\textbf{).}}
		\label{F}
	\end{figure*}
    \footnotetext[1]{The data is from BIS.}
    \footnotetext[2]{In trillions of US dollars, amounts denominated in currencies other than the US dollar are converted to US dollars at the exchange rate prevailing on the reference date.}
\section{Background}\label{sec2}

\subsection{Traditional Financial Infrastructure}
	Financial market infrastructure plays a critical role in the financial system and the broader economy. In the traditional framework of financial market infrastructure, functions such as securities registration, clearing, and settlement are provided by central institutions such as \textit{central securities depository, securities settlement system, central counterparty, payment system,} and \textit{transaction repository}. The central institutions record securities transactions and the balance of each account on the central server \cite{ref57}. 
	
	The central securities depository, securities settlement system, central counterparty, payment system, and transaction repository work together in the financial market to form an ecosystem of financial market infrastructure \cite{ref58}\cite{ref59}\cite{ref60}. We briefly describe these central institutions in Table \ref{F}. They each have different tasks and responsibilities while coordinating and supporting each other to ensure the efficient, stable, and secure operation of the financial market.
	
\begin{figure}
		\includegraphics[scale=0.45]{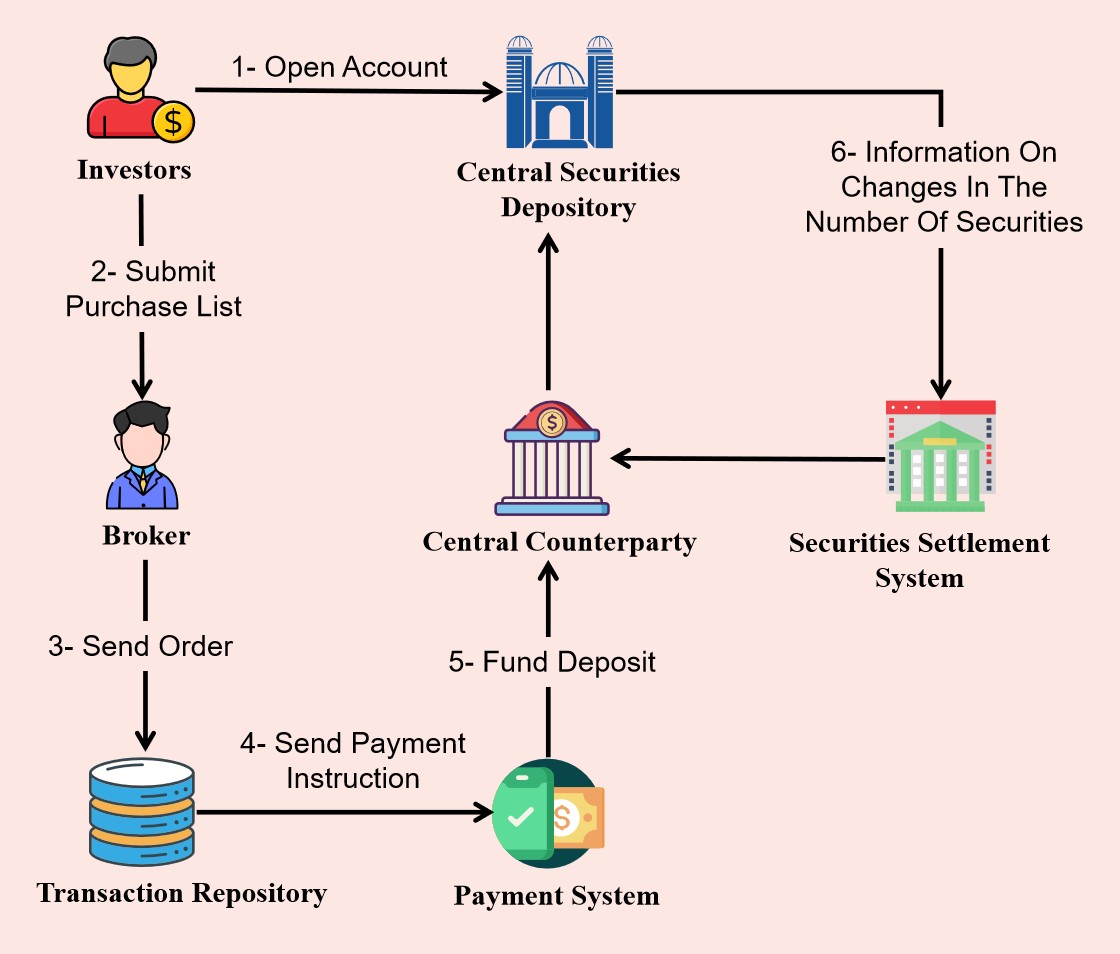}
		\centering
		\caption{\textbf{The process of collaborative work of financial institutions.}}
		\label{fig1}
	\end{figure}

    The traditional securities market is a centralized platform as shown in Fig.\ref{fig1}, which presents the workflow when a new investor participates in the platform. Firstly, investors need to open an investor account with the central securities depository (step 1) \cite{r3}. They also need to open a trading account with a broker (step 2). Once the investor's information is verified, they can buy or sell orders for stocks through the broker. The broker is responsible for submitting customer orders to the transaction repository (step 3). The transaction repository will validate and match the trade orders, and once matched, it will send a payment instruction to the payment system (step 4). The payment system will deduct the corresponding funds from the buyer's account and deposit them into the central bank (step 5). The central securities depository is responsible for processing the securities in the transaction, confirming the changes in the number of securities in the buyer's and seller's securities accounts. Then, the securities settlement system will transfer the ownership of the securities to the buyer's account (step 6). Finally, the central counterparty will settle with both the buyer and the seller, paying the net amount to the seller.
	
    Each of these steps requires communication and coordination between the \textit{central securities depository, securities settlement system, central counterparty, payment system,} and \textit{transaction repository}, ensuring that the transaction can be accomplished smoothly.

    However, the traditional financial infrastructure has limitations, which are listed below:

\begin{table*}[htb!]
    \renewcommand\arraystretch{1.5}
    \caption{\textbf{Description of various financial market infrastructure(FMI).}\newline}
    \label{F}
\begin{tabular}{|c|l|}
\hline
\textbf{Financial Infrastructure}            & \textbf{Description }                                                                                                                                                                                  \\ \hline
Payment System (PS)                 & A financial system that transfers monetary value from one entity to another                                                                                                               \\ \hline
Central Securities Depository (CSD) & A financial institution that manages investor accounts and post-trade cash settlements                                                                                                         \\ \hline
Securities Settlement System (SSS)  & A financial system that manages the settlement process for securities transactions                                                                                                             \\ \hline
Central Counterparty (CCP)          & \begin{tabular}[c]{@{}l@{}}A financial institution that acts as a counterparty between buyers and sellers and settles transactions \\ between parties involved in the transaction\end{tabular} \\ \hline
Trade Repository (TR)               & An entity that maintains a centralized electronic record (database) of transaction data                                                                                                      \\ \hline
\end{tabular}
	\end{table*}
   
    \begin{itemize}
		\item[	1)]\textbf{High transaction costs.} Traditional securities trading requires cumbersome steps and requires the involvement of third-party institutions. Consumers not only bear the trading risks during the transaction process but also need to pay corresponding commissions to the third-party institutions, which increases the overall transaction costs \cite{r6}\cite{r4}.  These costs ultimately reflect in customers' bank fees, interest rates, and exchange rates. In addition, the presence of multiple intermediary institutions in the trading chain means that each intermediary institution needs to maintain its trading system and records, resulting in asymmetrical and inconsistent trading information and further driving up transaction costs.
		\item[	2)]\textbf{Low liquidity.} Asset trading is subject to multiple restrictions, such as securities trading requiring confirmation and settlement by multiple intermediaries, and fund transfers requiring approval and settlement by financial institutions such as banks \cite{r6}. These restrictions result in a reduced liquidity of financial assets, making it difficult to complete transactions quickly and efficiently. This is particularly evident in cross-border transfers of funds, payments, and transfers, which typically require approval and settlement through multiple steps involving banks or other financial institutions.
		\item[	3)]\textbf{Low transparency.} Traditional financial infrastructure often incurs trust issues due to the lack of transparency regarding client assets \cite{r6}\cite{r3}. For instance, clients appoint securities companies as agents in securities trading but often lack a full understanding of the companies’ operational methods and strategies. Moreover, traditional financial institutions often do not provide clients with accurate and complete market information, which makes it hard for clients to make informed investment decisions. Furthermore, in traditional financial infrastructure, transactions and settlements usually require the participation of multiple financial institutions, each with its trading accounts and records. This can potentially lead to information asymmetry issues, resulting in errors or fraud during the transaction and settlement process, which in turn can lead to trust issues for customers \cite{r4}.
		\item[	4)]\textbf{Low security.} Currently, the number of participants in the securities market has reached 200 million, with daily trading volumes reaching hundreds of billions of yuan \cite{r1}\cite{r2}\cite{r3}. To reduce costs and improve security, a central counterparty is used to act as the trading counterparty for both buyers and sellers in securities settlements. However, However, this concentrates the risk on the central counterparty. Once a large-scale default occurs, the central counterparty will suffer an unbearable loss. At the same time, this poses a security threat on settlement data which is stored in centralized financial institutions. Once they are attacked, inevitably occurs data theft \cite{ref8}.
		\item[	5)]\textbf{Inefficient trading.} In traditional securities trading, investors are required to first open securities and fund accounts. Subsequently, investors prepare an order and submit it to a third-party institution, delegating them to buy or sell securities on their behalf. After submitting the order, investors need to wait for the third-party institution to accept the order and execute the transaction. Once the transaction is completed, the third-party institution performs clearing and settlement procedures to deliver funds and securities to the respective parties involved in the transaction. Furthermore, the settlement process involves multiple institutions with their responsibilities and requirements, resulting in a lengthy and cumbersome process with low efficiency \cite{ref9}.
	
	\end{itemize}

Financial applications are subject to specific regulatory requirements, aimed at ensuring the stability, reliability, and security of the financial system, such as:

\textbf{Anti-Money Laundering (AML)}: Financial applications are required to adhere to legal standards to mitigate and combat illicit activities, particularly in the realm of money laundering;

\textbf{Know Your Customer (KYC)}: Through a comprehensive understanding of clients, financial institutions are better equipped to identify anomalous activities, ensuring compliance with regulatory requirements and proactively preventing financial crimes.

\subsection{FinTech}
FinTech\cite{ref144} refers to the use of advanced technologies and innovative solutions to enhance and streamline financial services and business processes, thereby improving efficiency and effectively reducing costs. FinTech fosters the digital transformation of the financial industry. The technologies involved in FinTech exhibit characteristics of rapid iteration and cross-disciplinary nature. Common FinTech examples include artificial intelligence, big data, and blockchain.

\textbf{Artificial Intelligence}: Artificial intelligence plays a key role in financial technology \cite{ref145}. Through quantitative modeling, it helps financial institutions in risk management, credit assessment, and fraud detection, provides customers with personalized financial advice, and achieves cost reduction and efficiency improvement for financial institutions. Artificial intelligence accelerates the decision-making process by automating and intelligently processing tasks, thereby promoting the development of financial services.

\textbf{Big Data}: Big data analysis plays a key role in financial technology. We may use big data to analyze and extract valuable information for risk assessment and market trend analysis \cite{ref146}. By processing huge data sets, big data analysis provides more accurate risk assessment, more precise market trend forecasting, and more optimized investment strategies.

\subsection{Blockchain}
	Blockchain was originally designed to support the implementation of Bitcoin. In the Bitcoin system, all transactions are recorded in a public distributed ledger known as a blockchain. Each transaction in the ledger is encapsulated in a block and linked together using cryptographic methods to form an immutable chain. Each block contains the hash value of the previous block, which means that any attempt to tamper with a transaction record in any block will compromise the integrity of the entire blockchain. 
	
	Over time, the application of blockchain technology has expanded beyond its use in supporting Bitcoin, and it is now being widely utilized in various fields \cite{ref13}, particularly in the realm of financial technology, such as securities settlement \cite{ref147}, Cross-border payment \cite{ref148}, supply chain finance \cite{ref149}, Credit-Investigation System \cite{ref150}, and more. With extensive potential applications of blockchain technology, it is expected to be used in even more fields in the future.
	
 \begin{table*}[htb!]
	\renewcommand\arraystretch{1.5}
	\caption{\textbf{A comparison of three different types of blockchain systems.}\newline}
	\label{A}
	\centering
	\begin{tabular}{llll}
		\hline
		\makebox[0.25\textwidth][l]{Property}            & \makebox[0.25\textwidth][l]{Public}          & \makebox[0.25\textwidth][l]{Private}                  & \makebox[0.1\textwidth][l]{Alliance}            \\ \hline
		Consensus           & PoW \cite{ref22}, PoS \cite{ref23}          & PoA \cite{ref38}, BFT \cite{ref39}, FBFT \cite{ref94}             & PoA \cite{ref38}, BFT \cite{ref39}               \\
		Mechanism           & All miners       & Centralised organisation & Leader node set       \\
		Protocol Efficiency & Low efficiency   & High efficiency          & High efficiency       \\
		Consumption         & High energy      & Low energy               & Low energy            \\
		Management          & Permissionless   & Permissioned whitelist   & Permissioned nodes    \\
		processing speed    & Order of minutes & Order of milliseconds    & Order of milliseconds \\ \hline
	\end{tabular}
\end{table*}
	
	\subsubsection{Principles of Blockchain Technology}
	\ 
	\newline 
	 \newline The maintenance of the shared ledger among distributed nodes can be reduced to a mathematical problem known as the Byzantine Generals' Problem \cite{ref17}, which is used to avoid being deceived and confused by malicious attackers when people need to exchange value with unfamiliar opponents \cite{ref18}. In the technical field, the Byzantine Generals' Problem can be summarized as how each node in a network can reach a consensus without a trusted central node and trusted channel. Blockchain technology solves the well-known Byzantine problem by proof-of-work(POW) \cite{ref22} or proof-of-stake(POS) \cite{ref23}. To ensure the safety of the ledger, blockchain combines \textit{distributed storage, consensus mechanism,} and \textit{cryptography technology}. To automatically enforce contract terms, most blockchains incorporate smart contracts \cite{ref19}.     
	 \newline \indent \textbf{Distributed storage:} In blockchain, data storage is not handled by a central node, but rather by all nodes on the network working together. Each node maintains a complete copy of the ledger, meaning that even if a node fails, data can still be retrieved from other nodes. This distributed storage approach ensures data security and reliability to a greater degree \cite{ref20}.
	 \newline \indent \textbf{Consensus mechanism:} Consensus algorithms refer to how nodes in a distributed network make agreed-upon decisions \cite{ref21}. Because blockchain data storage is distributed, consensus must be reached between each node to ensure that all ledger copies are the same. Common consensus mechanisms include proof-of-work \cite{ref22}, proof-of-stake \cite{ref23}, and others. In the proof-of-work mechanism, nodes need to complete certain computational tasks to gain the right to record transactions, thereby ensuring the credibility of data, whereas, in the proof-of-stake mechanism, nodes need to possess a certain amount of digital assets to gain the right to record transactions.
	 \newline \indent \textbf{Cryptography technology:} Blockchain uses cryptography technology to ensure the security and privacy of data. What we are mainly concerned with here are hash functions and public-private key encryption algorithms. A hash function is a one-way function that can convert arbitrary-length data into a fixed-length hash value, and it is impossible to reverse the hash value to the original data. Public-private key encryption algorithm refers to the use of a pair of keys, one of which is a publicly available public key, and the other is a private key that is kept secret. Data encrypted with a public key can only be decrypted using the corresponding private key, ensuring the confidentiality of data \cite{ref24}.
     \newline \indent \textbf{Smart contract:} Smart contract is a set of commitments defined in digital form, including agreements on which contract participants can execute these commitments \cite{r6}. They aim to automate the execution of contracts, thereby eliminating intermediaries and reducing transaction costs \cite{ref90}. Smart contracts can be programmed to implement various conditions and constraints, such as automatically executing payments based on specific events or times, checking account balances, creating digital assets, and more. These program codes are embedded into the blockchain network and therefore cannot be tampered with when executed on the network. Smart contracts are widely used in finance, insurance, supply chain, real estate, and other fields to achieve secure, transparent, and efficient business processes \cite{ref91}.

	\subsubsection{Blockchain System Classification}
	\ 
	\newline 
	\newline Based on the degree of centralized control, blockchain can be grouped into \textit{public chain, alliance chain,} and \textit{private chain} \cite{ref32}. These three types of blockchains have their own advantages and disadvantages, which can be summarized as below:
        \newline \indent \textbf{Public Blockchain:} The public blockchain is an open, transparent, and decentralized network that is accessible to all blockchain service clients. This high degree of transparency and trustworthiness is a key feature of the public blockchain \cite{ref92}\cite{ref93}. All nodes that participate can collaboratively verify, record, and store transaction information. The data on the public blockchain is publicly visible, allowing anyone to view and verify its authenticity. 
        \newline \indent \textbf{Alliance Blockchain:} An alliance blockchain is a controlled network consisting of specific members \cite{ref33}. The participants are restricted and certified entities or organizations who establish trust relationships by jointly maintaining and verifying transaction information \cite{ref34}. Compared to public blockchains, alliance blockchains are more flexible as they can be designed and managed according to specific needs and rules. They can provide higher transaction throughput, faster confirmation speeds, as well as greater privacy protection and data control based on specific requirements. Participants in an alliance blockchain typically sign agreements that specify how the ledger is managed and used, and they share access and management rights. This model provides higher trust and compliance while offering participants more flexibility and control \cite{ref35}.
        \newline \indent \textbf{Private Blockchain:} A private blockchain is a network that is exclusively controlled and managed by a particular entity or organization. Only authorized nodes can access it, and data can be read and modified according to predefined rules. Private blockchains ensure high security and privacy protection as all data is private and only accessible to authorized nodes. Additionally, they offer high performance and throughput since they do not require a consensus algorithm or mining process. Validation and confirmation of transaction information is limited to authorized users \cite{ref36}\cite{ref43}.     
	\newline \indent Table \ref{A} provides a comparative summary of key characteristics of the three types of blockchains. Due to variations in technical solutions, the application scenarios for public, private, and alliance blockchains also differ. Public blockchains are majorly used in social life and modern business fields \cite{ref42}. Private blockchains are primarily used for internal work processes such as enterprise database management and auditing, which can also be applied in government scenarios \cite{ref44}. The alliance chain is mainly a specific application between institutions that can be used in supply chain finance, electronic forensics, and other businesses \cite{ref33}\cite{ref34}\cite{ref45}. Due to the strict control of transactions and information confidentiality in the financial field, as well as suitability requirements for enterprise participants, alliance blockchains are more appropriate for the securities industry.  

\begin{table*}[htb!]
\renewcommand\arraystretch{1.5}
\caption{\textbf{A comparative of IPO, ICO, and STO.}\newline}
\label{B}
\centering
\begin{tabular}{llll}
\hline
       \makebox[0.2\textwidth][l]{Property} & \makebox[0.25\textwidth][l]{IPO}                                                                                & \makebox[0.25\textwidth][l]{ICO}                                                                         & \makebox[0.2\textwidth][l]{STO}                                                                       \\ \hline
Supervision                                                                         & High                                                                              & Low                                                                        & Medium                                                                    \\
Underlying assets                                                                   & Equity                                                                   & Right to use                                                     & Income rights                                               \\
Issue difficulty                                                                    & High                                                                               & Low                                                                         & Medium                                                                    \\
Transaction convenience  & Low & High & High \\   Transaction security  & High & Low & Medium \\
Investment threshold                                                                & High                                                                               & Low                                                                         & Medium          \\ 
Period                                                                & Years or months \cite{ref175}\cite{ref176}                                                                                & Weeks \cite{ref175}\cite{ref176}                                                                        & Months \cite{ref175}\cite{ref176}\\ \hline                                                       
\end{tabular}
\end{table*}

	\subsubsection{Features of Blockchain Technology}
	\ 
	\newline 
	\newline \indent \textbf{Decentralization}. Decentralization is a key characteristic of blockchain technology \cite{ref25}. Unlike traditional systems that rely on centralized authorities or intermediaries to validate transactions and maintain records, blockchain networks distribute these functions among a large network of nodes. Each node has equal rights to record and verify data. In the real world, trust is often established through the use of third-party intermediaries. These intermediaries provide trust endorsements and help to resolve disputes. However, the architecture of these third-party service providers is typically private and centralized, leading to issues with transparency, accountability, and security. Blockchain technology uses a decentralized database architecture to address these issues. By using a distributed network of nodes, blockchain networks can establish algorithmic trust and eliminate the need for centralized trust models. This opens up new possibilities for global mutual trust and connectivity, as it allows for secure and transparent exchange of data and assets across borders \cite{ref26}.
	 \newline \indent \textbf{Immutability} \cite{ref27}. Once data is added to the blockchain, it is permanently stored and cannot be modified. In the structure of a blockchain, each block contains a unique hash value of its previous block which can be verified but is extremely difficult to crack. In this way, blocks containing information are linked together to form a main chain, which is then stored in a distributed manner across blockchain nodes. Unless an attacker can control more than 51$\%$ of nodes in the system \cite{ref28}, any attempt to modify the data on a single node will be rejected by the rest of the network. Therefore, the data stored on a blockchain is considered to be highly stable and reliable.
	 \newline \indent \textbf{Transparency and anonymity} \cite{ref29}. Data in a blockchain is stored in a decentralized manner across all nodes in the network, and each node can view all transaction information. Additionally, all nodes are updated simultaneously as new transactions are added to the network. Because the blockchain is based on purely mathematical principles, which adopt public key addresses rather than the personal information of the transacting parties. This provides anonymity within the framework of transparency, meaning that the data recorded on the blockchain is both transparent and open, yet anonymous and reliable.
	  \newline \indent \textbf{High availability}. Blockchain technology adopts a distributed computing model, which distributes data and computing resources across multiple nodes in the network. This means that even if a node is attacked or occurs failure, the entire system is still capable of normal running to ensure high availability \cite{ref30}. Additionally, as data is distributed across different nodes, the system's performance is more stable and less susceptible to failures or outages.
	\newline \indent \textbf{Efficiency}. Blockchain technology is highly efficient by adopting smart contracts, which can automate execution and avoid the intermediate links in traditional transactions. In this way, it reduces transaction costs and improves efficiency \cite{ref31}. Additionally, the automation and intelligence of the blockchain system realize unmanned transactions, avoiding human intervention and errors in traditional transactions, thereby improving the reliability and accuracy of transactions.
\ 
\

\section{Blockchain Platforms Deployed in Finance Areas}

With the continuous development of blockchain technology in the financial field, many blockchain platforms have emerged, including Corda \cite{ref106}, Quorum \cite{ref114}, and tZERO \cite{ref100}. These platforms are designed to provide secure, fast, low-cost, and reliable financial transactions while ensuring the traceability and immutability of transactions. They typically employ cryptographic techniques and consensus protocols to achieve transaction security and transparency. Due to these characteristics, these platforms have achieved a wide range of application scenarios. The Defi ecosystem is an important product of this process. Financial blockchain serves as the underlying infrastructure, while Decentralized Finance (DeFi) constitutes applications built upon this foundational framework. Financial blockchain contributes technical support and infrastructure for DeFi, enabling the feasibility of decentralized financial systems.

 However, each platform has many different constraints in practical applications. Therefore, it is essential to thoroughly consider factors, such as their performance and reliability, before we decide to use these platforms. This section will delve into six mainstream blockchain implementations.

\subsection{Ethereum}
Ethereum is a decentralized open-source blockchain platform that enables developers to create decentralized applications (DApps) using smart contracts \cite{ref104}. The primary objective of Ethereum is to establish a global, free, transparent, and tamper-proof infrastructure that allows people to freely create and use applications without relying on centralized institutions \cite{ref105}. Ethereum's unique features have made it one of the most influential blockchain platforms in the financial sector, receiving significant support and being applied in leading financial institutions and companies such as Microsoft, JPMorgan Chase, Accenture, ING, Intel, and Cisco, among others \cite{ref106}.

As one of the core technologies of Ethereum, smart contracts enable Ethereum to support a range of decentralized applications, including digital currency transactions, issuance, and trading of financial derivatives, among others. In \cite{r4}, the authors tackle the shortcomings of traditional centralized stock exchange systems(such as high transaction fees), by implementing a prototype in Ethereum for a subset of rules for the Bucharest Stock Exchange. At present, DeFi is mainly active within the Ethereum network ecosystem, attributing to various emerging financial innovation applications, including stablecoins, lending platforms, derivatives, prediction markets, insurance, payment platforms, and more.

Overall, Ethereum provides a robust infrastructure for the development of decentralized applications, making it one of the most significant platforms in the blockchain ecosystem. However, Ethereum's adoption of the PoW consensus mechanism requires substantial computing resources and energy consumption, leading to relatively low transaction speed and throughput, with only about 40 transactions processed per second \cite{ref106}. This limitation makes it challenging to meet the demands of large-scale financial transactions. Moreover, vulnerabilities of smart contracts have resulted in severe security issues along with asset losses. For instance, the DAO incident resulted from the reentrancy vulnerability of a smart contract that led to the theft of approximately $\$$50 million worth of ETH \cite{ref107}.

\subsection{Hyperledger Fabric}

Hyperledger Fabric is an open-source distributed ledger technology platform hosted by the Linux Foundation, which has received widespread support from numerous enterprises and organizations \cite{ref106}. Compared to other public blockchains, Hyperledger Fabric is a permissioned private blockchain platform, meaning that participants in the Hyperledger Fabric network must be authorized to join. This platform is primarily used for enterprise internal applications to ensure data security and privacy. As a trusted, secure, and efficient enterprise-grade blockchain platform, Hyperledger Fabric can provide secure, fast, low-cost, and reliable transaction solutions for the financial industry \cite{ref108}.

In the financial industry, Hyperledger Fabric has been applied in various aspects, such as asset management, supply chain finance, risk management, settlement, and clearing of securities transactions. One example of this is We. Trade, which is the world’s first enterprise-grade, blockchain-enabled trade finance platform that offers a safe and more reliable platform for buyers and sellers to trade globally using distributed ledger technology and smart contracts \cite{ref112}.

The advantage of Hyperledger Fabric lies in its modular and multifunctional design, which caters to a wide range of industry use cases. It allows for plug-and-play components, and its unique consensus approach enables privacy protection while achieving scalable performance. Additionally, Hyperledger Fabric leverages Kafka \cite{ref124} to sort and process transaction information, which provides high throughput and low latency processing capabilities, and supports node fault tolerance within the cluster. Compared to Ethereum, Hyperledger Fabric processes transactions much faster \cite{ref106}. However, Hyperledger Fabric also bears some challenges and limitations in practical applications, one of which is the possible lack of transparency. This is because Hyperledger Fabric is a permissioned private blockchain platform, allowing participants to have specific access rights and control in the blockchain network. Therefore, in certain situations, some participants in the network may have excessive control permissions, which could pose a secure risk of data monopolization or tampering.

\subsection{Corda}
Corda is an open-source software developed and launched by R3 blockchain technology company. Tailored specifically for commercial applications, Corda stands as a permissioned blockchain platform \cite{ref106}. In the realm of the financial sector, Corda showcases its versatility across diverse domains, encompassing trade financing, securities issuance, asset management, insurance, and interbank payments, among other crucial applications. Noteworthy is the widespread adoption of Corda by financial institutions worldwide, exemplified by its integration into the Nasdaq stock exchange for bolstering market infrastructure, and its active role within Italian banking institutions for facilitating interbank payments \cite{ref113}. One of the core strengths of Corda lies in its capacity to ensure scalability while simultaneously accommodating decentralized assets. It achieves this while upholding stringent privacy standards and regulatory compliance. This intricate blend of attributes positions Corda as a choice for institutions seeking to navigate the intricate landscape of modern finance. 

Corda is a decentralized network, comprising an assemblage of Corda nodes. This framework empowers participants to engage in transactions and information sharing. Similar to Ethereum, Corda utilizes smart contract technology to automate and digitize business processes \cite{ref109}. However, Corda requires consensus solely among the participating parties, which significantly augments transaction efficiency and scalability. It is noteworthy that transaction details of corda remain exclusively visible to the participants involved, thereby safeguarding transaction privacy and the security of crucial business secrets. Comparing Corda to other blockchain platforms, it exhibits a higher transaction rate than Ethereum, although its throughput falls slightly behind that of Hyperledger Fabric \cite{ref106}.

\subsection{Quorum}

Quorum is an open-source distributed ledger platform based on Ethereum technology, which is a fork of the Ethereum Go language implementation version. It leverages a voting-based consensus algorithm. The unique highlight of Quorum, which ensures data privacy is the new feature known as a private transaction identifier. This feature allows transaction senders to create private transactions by marking who is privy to a transaction via the \textit{privateFor} parameter. Private data is stored off-chain in a separate component called the private transaction manager, which encrypts private data, distributes the encrypted data to other parties that are privy to the transaction, and returns the decrypted payload to those parties. This helps ensure that only the intended recipients can view the contents of a private transaction. The primary goal of Quorum is to fully reuse existing Ethereum technology as much as possible. As a result, quorum blockchain use cases would have to undergo limited changes to maintain sync with upcoming versions of the public Ethereum codebase \cite{ref114}.

The most promising sector Quorum blockchain points out to financial services. The most prominent applications of quorum blockchain in finance include tokenized cash, commercial bank payments, trade finance, supply chain finance, institutional trading, capital market data, commodity post-trade processing, loan marketplaces, and interbank payments in association with central banks. Apart from these applications, Quorum is also applied to develop a ledger system for auditing financial transactions. For example, Block Ledger is a well-known application of Quorum. It is a decentralized accounting ledger system that leverages Quorum through the BaaS (Blockchain-as-a-Service) approach. The functions of the Block Ledger focus on rationalizing debtors and creditors through the addition of hash on the blockchain. This is very beneficial for account reconciliation, transparency, risk and credit scoring, e-invoicing, and audit trail \cite{ref114}. 

To ensure high throughput and fast transaction confirmation, Quorum adopts the Raft consensus algorithm. Unlike Ethereum, Quorum supports private transactions and can be configured as a private network that only allows specific participants to transact. Additionally, Quorum supports private deployment of smart contract code, which can help enterprises protect their smart contract code and data from being publicized. Furthermore, Quorum provides many extensions and additional features such as enterprise-grade identity authentication, data privacy, and integrated APIs. The platform also supports interoperability with other blockchains and traditional financial systems, providing more possibilities and opportunities for financial services. Quorum aims to be an efficient, secure, and scalable distributed ledger platform suitable for the financial sector \cite{ref110}.

\subsection{Symbiont Assembly™}

Symbiont Assembly™ is a blockchain platform developed by Symbiont. Unlike public blockchains, Symbiont Assembly™ is a permissioned blockchain, of which network access is restricted to authorized participants. Symbiont Assembly™ aims to provide a secure, efficient, and transparent distributed ledger solution for enterprises and financial markets. It uses blockchain technology to ensure data security and immutability, and offers smart contracts to functionally automate and simplify transaction processes \cite{ref103}.

Specifically, Symbiont Assembly™ supports the issuance and trading of securities, bonds, and other financial assets. On the Symbiont Assembly™ platform, issuers utilize smart contracts to define characteristics of securities or bonds, including issuance volume, circulation time, yield, etc. These smart contracts will be encoded on the blockchain, ensuring that their execution can not be tampered with by anyone. In addition, investors can purchase, hold, and trade these securities or bonds via the platform which provides higher liquidity for the market.

\begin{figure*}[htb]
	\includegraphics[scale=0.6]{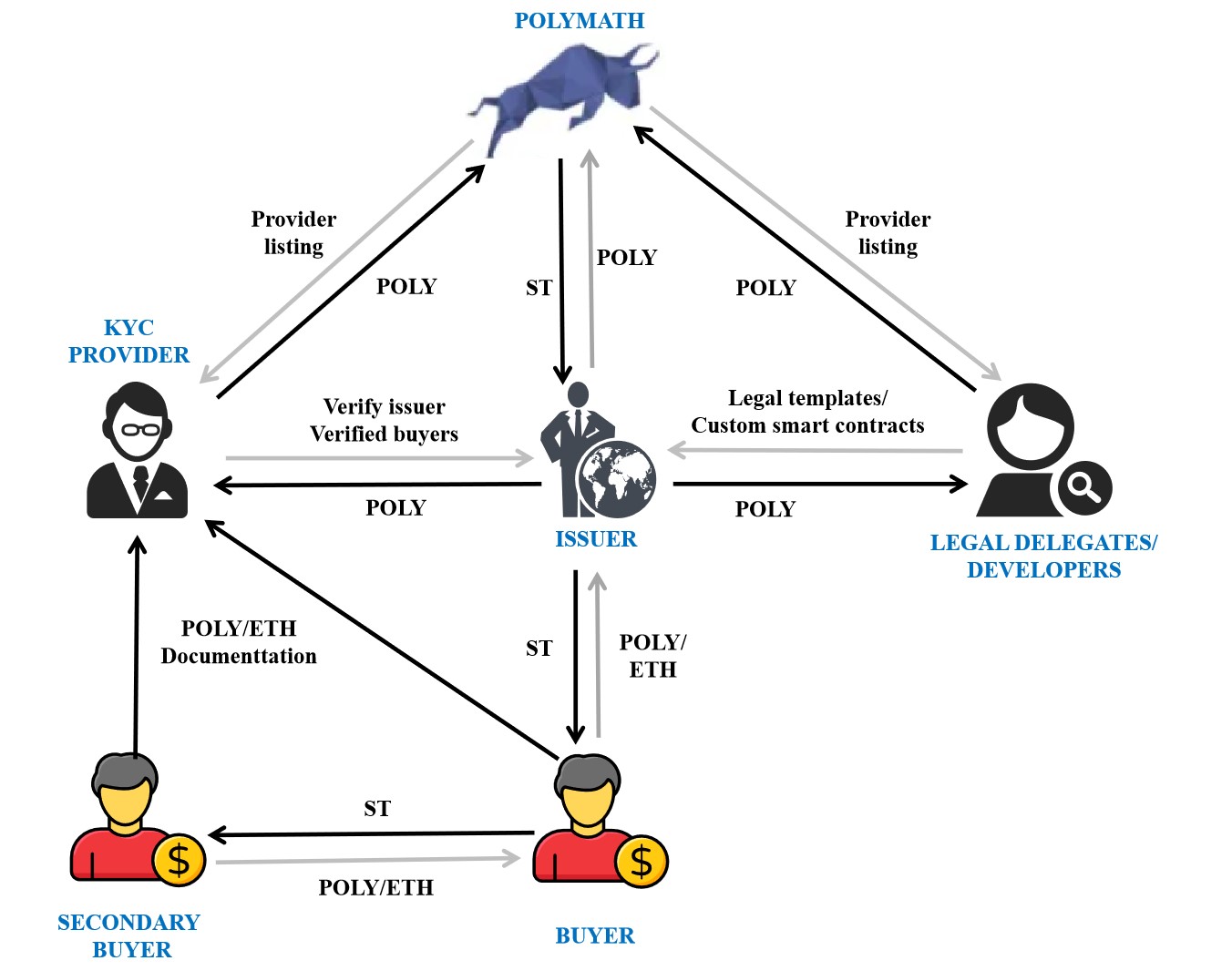}
	\centering
	\caption{\textbf{The security token issuance and transaction process on the Polymath platform.}}
	\label{fig3}
\end{figure*}

\subsection{tZERO}

tZERO is a blockchain trading platform aiming at providing investors with faster, more transparent, and more secure securities trading. tZERO simplifies securities trading and reduces transaction costs with blockchain technology which is able to eliminate the complex network between intermediaries and exchanges \cite{ref100}.

tZERO is a regulated and licensed platform that supports the trading of traditional private securities and blockchain-based digital securities, including conventional security tokens and non-fungible tokens (NFTs). By using smart contracts, the platform automates many tasks involved in traditional financial transactions, such as securities trading settlement and asset management. These automated features not only improve transaction efficiency but also reduce transaction costs. The tZERO platform also features highly secure characteristics by utilizing Blockchain technology, such as the use of multi-signature and distributed storage technologies to protect users’ digital assets and transaction data. Additionally, the tZERO platform uses digital identity verification and smart contracts to ensure the identity and compliance of the trading parties. Finally, tZERO has developed a token called TZROP, which is used to purchase securities and trading services on the tZERO platform. The use of this token brings more liquidity to the platform and higher trading efficiency of participants.

\section{Blockchain Applications in Finance}\label{sec3}

Blockchain applications in finance can be roughly classified into four categories, namely \textit{capital raising, securities trading, financial analysis,} and \textit{investment management}. In what follows, we will elaborate on each of the four categories one by one.

\begin{figure}
	\includegraphics[scale=0.55]{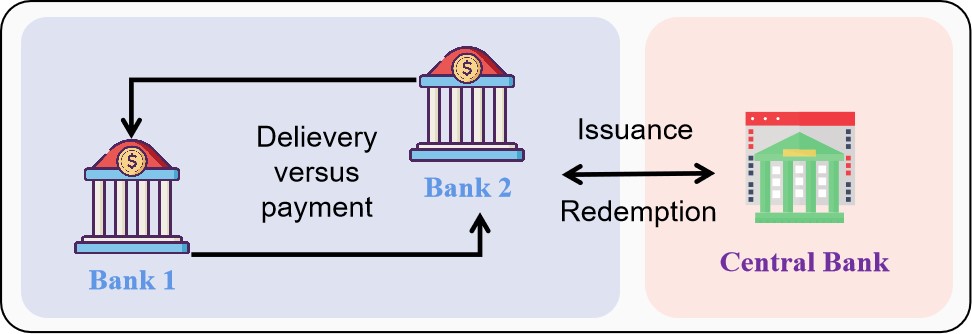}
	\centering
	\caption{\textbf{The two-tier structure.}}
	\label{fig14}
\end{figure}
\subsection{Capital Raising}

The healthy operation of the securities market requires sufficient capital supply. On the contrary, securities issuance is also an efficient way for different enterprises to engage in direct capital raising. However, capital raising by issuing securities has problems, such as information asymmetry, lack of trust, complicated transaction process, and low liquidity \cite{r6}. 

The blockchain has unique characteristics, such as decentralization, high transparency, enhanced security, and immutability of information, making it a good solution to the above problems \cite{r6}. Over the past few years, blockchain technology has been extensively utilized in the field of securities investment, giving rise to a novel financing mechanism called tokenized securities or security token offerings (STOs) \cite{ref95}. By leveraging blockchain technology, this mechanism merges traditional securities with digital currencies, thereby providing investors and issuers with a more adaptable, transparent, and efficient way to raise capital.

STO is an attempt by governments, such as the United States government, to bring the existing Initial Coin Offering(ICO) market into a traditional financial infrastructure framework without enacting new regulatory policies. The ICO market is growing rapidly and has the characteristics of internationalization, decentralization, and lack of regulation. However, as a widely targeted and completely uncontrolled financial product, ICO is a headache for regulatory agencies in various countries. To overcome the ICO’s lack of regulation and transparency, STO combines the characteristics of ICO and Initial Public Offering (IPO). In Table \ref{B}, we provide a brief comparison of IPO, ICO, and STO.
\begin{figure*}[htb]
	\includegraphics[scale=0.5]{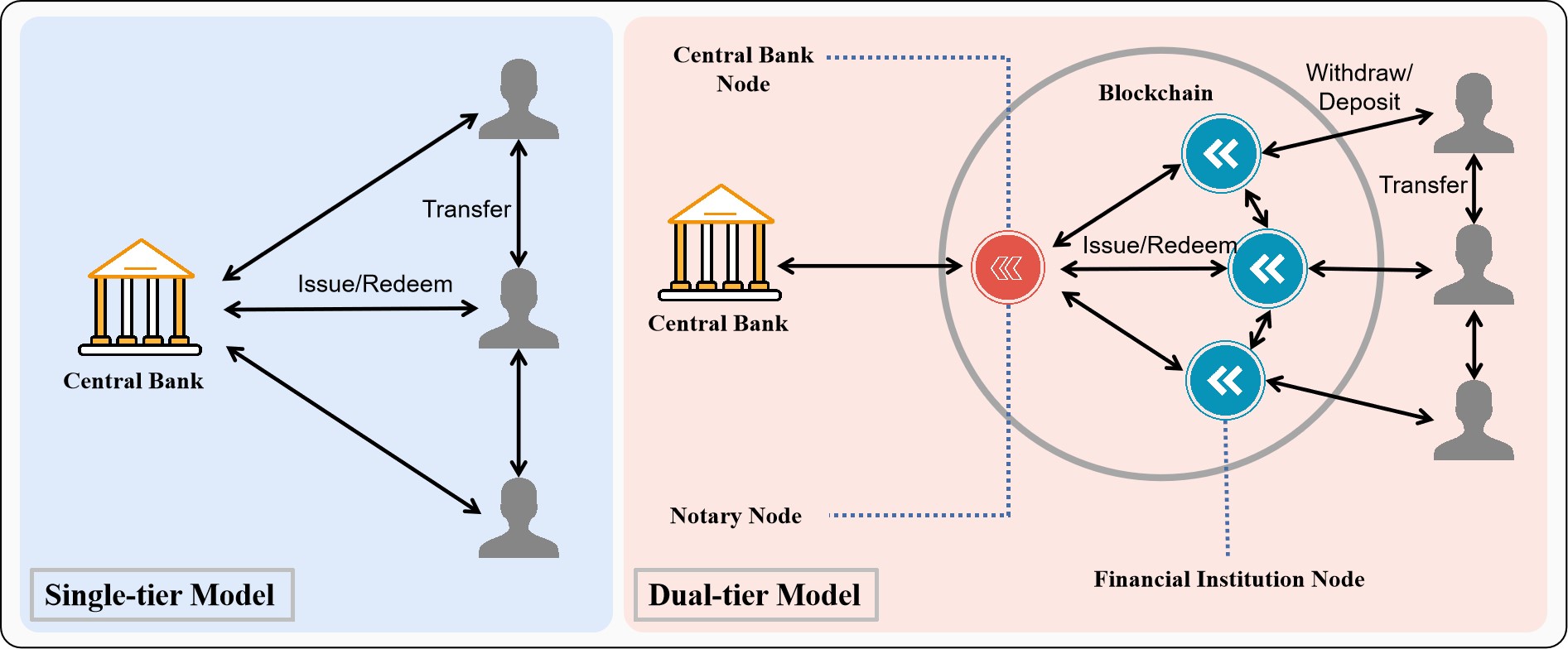}
	\centering
	\caption{\textbf{The two models of retail CBDC.}}
	\label{fig15}
\end{figure*}

The STO issuance platform is a vital component of the new financing model, of which the technical principle is to apply blockchain technology and smart contracts to realize the issuance, management, and trading of security tokens. On these platforms, issuers can conduct compliant token issuance for investors to buy and trade tokens, including various types of securities such as equity, debt, and funds. STO tokens are compliant with regulations, offering investors higher security and protection. Currently, there are many STO issuance platforms available, such as Polymath \cite{ref97}, Securitize \cite{ref98}, Harbor \cite{ref99}, tZERO \cite{ref100}, and TokenSoft \cite{ref101}. These platforms explore blockchain technology to ensure the security and transparency of tokens while providing compliant token issuance and trading services. They aim to help issuers and investors achieve better capital raising and profit-making.

As a case in point, consider Polymath \cite{ref97}  which exploits the Ethereum blockchain as the basis for its token issuance and applies smart contracts to create and manage security tokens. The issuance process of Polymath platform security tokens is shown in Fig.\ref{fig3}. Firstly, Polymath creates a non-active security token for the issuer, which cannot be traded until it is signed by the designated authorized representative. Secondly, the legal representative reviews the issuance details, improves the proposal with detailed information on the offer for issuance, and provides price suggestions for the issuer to choose. After completing the compliance process, the legal representative sets the initial issuance contract address. Once the issuer is ready, trading will start up, the procedure of which is specified by the smart contract of security token issuance. Investors choose a KYC (Know Your Customer) provider and pay POLY(Digital Currency on Polymath) to be added to the investment whitelist. KYC's provider is responsible for reviewing investor identities. If everything is normal, the Ethereum address of the investor will be set as the white list and associated with their real identity. After the initial sale, investors can sell the token to second-tier investors who are listed in the KYC provider's whitelist.

Overall, STO is a financing approach based on digital securities, which has been realized through blockchain technology. This decentralized financing model is distinguished by the characteristics of transparency, security, and low cost, which offers investors and funders a financing approach that is more flexible, efficient, and convenient. The public and immutable features of blockchain technology enhance the transparency and reliability of STO transaction records, while the use of cryptographic algorithms ensures the security of digital securities, effectively preventing data tampering and data forgery. Furthermore, blockchain technology can reduce the issuance cost of digital securities, while increasing the liquidity of securities, which in turn prompts the growth of their value. In conclusion, by utilizing blockchain technology, STO offers a more efficient, secure, and cost-effective financing approach to the market, creating more earnings for investors and funders.

\ 

\subsection{Securities Trading}

The securities industry seeks to create and maintain an orderly securities market. Traditional securities markets rely on the mutual cooperation of financial infrastructures, such as central securities depositories, securities settlement systems, central counterparties, payment systems, and trading databases. However, the traditional securities market has inherent defects including \emph{high transaction costs, low liquidity, low transparency, inefficient trading, etc}.

To overcome the above issues, blockchain technology offers a decentralized distributed database system that enables trading participants to trade without intermediaries, leading to more efficient, transparent, and secure transactions. The introduction of blockchain technology into the securities market is causing a disruptive transformation, leading to the creation of blockchain-based financial infrastructures. In particular, decentralized payment methods based on blockchain allow transactions to occur outside of traditional payment institutions, which reduces both the transaction cost and secure risks. For example, Central Bank Digital Currency issued by central banks of various countries. Decentralized securities exchanges enable participants to trade directly, avoiding extra trading costs (Such as Nasdaq \cite{ref69}, Bucharest \cite{r4}, etc.). The blockchain-based clearing and settlement systems achieve openness, transparency, and reliability of transaction information, ensuring the security of transactions (Such as SETLcoin \cite{ref71}\cite{ref70}).

These blockchain-based financial infrastructures are promoting the development of the securities market towards digitization and intelligence, which bring both new opportunities and challenges to the securities market. Overall, the integration of blockchain technology into the securities market has revolutionized the way securities are traded, settled, and cleared, creating a more efficient and transparent market for all participants.

	
		

\
\ 
\ 
\subsubsection{Payment System in the Securities Industry}
\ 
\newline 
\newline A blockchain-based payment system is a digital currency implemented using distributed ledger technology, which can provide fully decentralized, fast, and transparent cross-border payment services. However, this payment system also suffers a  great number of challenges, such as insufficient supervision, high volatility, low scalability, etc. To address these issues, a few central banks are exploring issuing central bank digital currencies (CBDC).

CBDC is a digital currency issued and managed by a national central bank, which attempts to enhance the security and efficiency of monetary policy, financial stability, and payment systems. It utilizes blockchain and other distributed ledger technologies to replace existing payment methods in the real world. In response to the emerging digital payment environment, worldwide central banks are exploring the study and implementation of CBDC. According to the 2021 survey results of the Bank for International Settlements (BIS), at least 86$\%$ of the world's central banks have initiated relevant research on CDBC, and some countries have already run the testing phase, such as China, Sweden, and South Korea \cite{ref82}.

According to the definition provided by BIS, CBDCs can be categorized into two types based on the intended users: wholesale CBDCs primarily issued to large financial institutions such as commercial banks, and retail CBDCs primarily for consumers and businesses \cite{ref111}.

Wholesale CBDCs build on the current dual-tier structure(see Fig.\ref{fig14}), which places the central bank at the foundation of the payment system while assigning customer-facing activities to commercial banks and other payment service providers(PSPs). Wholesale CBDC is designed to settle interbank transfers and related transactions, such as payments between financial institutions. One significant advantage of wholesale CBDC settlement is that it enables new forms of conditional payments, which require settlement only upon the delivery of another payment or asset. 
These conditional payment instructions far exceed the delivery versus payment mechanism in today’s real-time gross settlement systems (RTGSs). Instead, wholesale CBDCs make central bank currency programmable to support automation[14]. 
Wholesale CBDCs typically exhibit higher transaction speeds and lower transaction costs, revealing the promising future for cross-border payments and large-scale transactions. 
Additionally, wholesale CBDCs enhance the security and transparency of the financial system as transactions can be traced and recorded, significantly reducing the risks of money laundering and other illicit activities.

Retail CBDC refers to digital currencies issued by central banks for use by the general public. Currently, there are two models of retail CBDC: single-tier model and dual-tier model(see Fig.\ref{fig15}). In the single-tier model, in addition to issuing CBDC, the central bank directly operates the payment system and provides retail services. In this way, all operations are maintained by the central bank. In the dual-tier model, CBDC is still issued by the central bank, but payment services and account maintenance are provided by large financial institutions such as commercial banks.

As the safest digital asset, CBDC will inject new vitality into the global monetary and financial system, becoming a new cornerstone for future payment transactions \cite{ref82}\cite{ref83}\cite{ref84}. CBDC inherits the public's trust in fiat currency and effectively fills the gaps in private digital currencies. Additionally, CBDC can improve payment efficiency by enabling near real-time transactions, lowering payment costs, and reducing intermediaries' involvement. CBDC also promotes financial inclusion, allowing more people to participate in the economy, especially by making cross-border payments more convenient and financial services more accessible. Moreover, CBDC features smart contract settlement functions, making currency transactions more intelligent by potentially establishing predefined standards that will automatically execute when transactions meet these standards. Finally, CBDC can strengthen the management of monetary policy as the central bank may better regulate the money supply and market liquidity by controlling CBDC issuance, achieving macroeconomic regulation objectives. However, since the implementation of CBDC is still in the early stages, some unknown issues may arise. Therefore, further research is needed to be carried out for more information on the performance of CBDC.

\begin{figure}
	\includegraphics[scale=0.6]{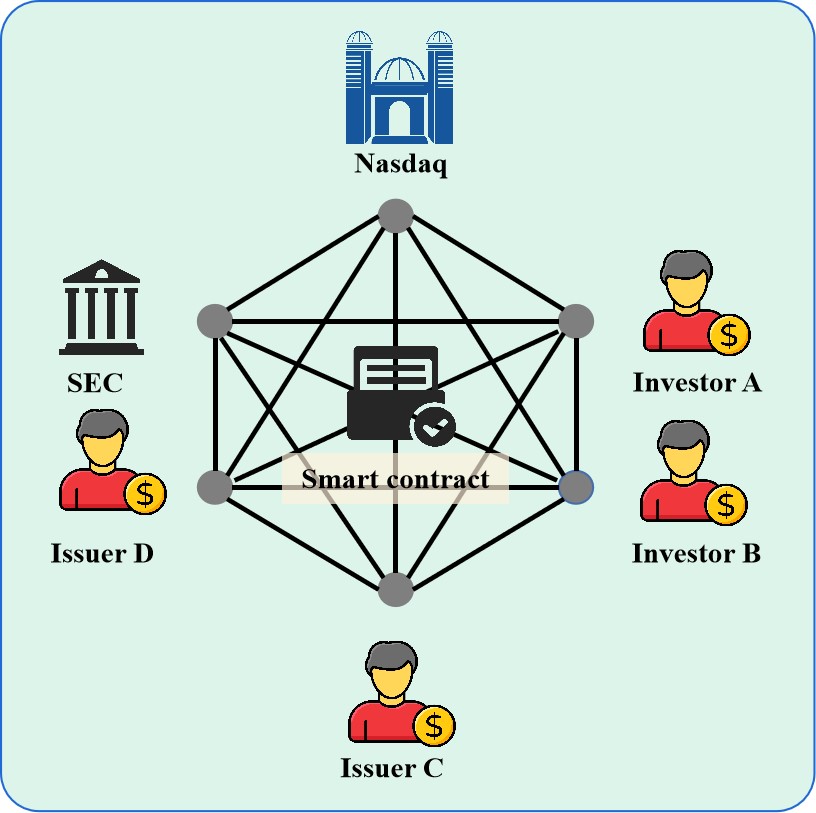}
	\centering
	\caption{\textbf{Nasdaq Linq platform  architecture.}}
	\label{fig8}
\end{figure}
\subsubsection{Trading System in the Securities Industry}
\ 
\newline 
\newline A securities trading platform is an electronic platform specifically designed for buying and selling stocks and other securities products. It allows investors to execute buy and sell transactions in the stock market and provides many tools and features to assist them in trading and managing their investment portfolios. Securities trading platforms are typically developed and provided by securities brokerage firms, investment banks, or financial technology companies. However, traditional securities trading platforms have problems such as low transaction efficiency and poor transparency. With the development of blockchain technology, blockchain-based securities trading platforms are gradually gaining attention and interest. These platforms utilize a decentralized architecture to achieve fast transactions. Additionally, blockchain technology offers improved transparency in trading.

Nasdaq Linq \cite{ref68} is a blockchain-based securities trading platform launched by Nasdaq in 2015, which improves the efficiency and transparency of securities trading. Specifically, Nasdaq Linq uses blockchain technology to record transaction information, including securities issuance and transfers. By utilizing blockchain technology, transaction information can be encrypted and stored on an immutable distributed ledger, enhancing transaction transparency and security. Additionally, the platform is capable of automated settlement and clearing functions, which further increases transaction efficiency \cite{ref69}.

As shown in Fig.\ref{fig8}, Nasdaq Linq is a private blockchain that does not need proof-of-work or other consensus mechanisms. Participating nodes include Nasdaq, SEC, Issuers, and Investor. In this platform, Nasdaq acts as a trusted intermediary to run and monitor the blockchain. SEC is a regulatory agency in the United States responsible for enforcing federal securities laws and regulating the securities industry. The shares issued on Nasdaq Linq must follow the rules of the U.S. Securities and Exchange Commission (SEC). Issuers must file some documents with the SEC, offering the basic information of the issuance. Only authorized participants can access and verify transactions. Nasdaq Linq uses smart contracts to make private equity management and regulation easier and to enable functions like automatic issuance, transfer, and dividend distribution of equity.

The introduction of blockchain technology in the Nasdaq Linq platform has brought a great deal of profits. In general, equity financing and transfer transactions for unlisted companies required a lot of manual labor and paper-based work, involving manual handling of paper stock certificates, option grants, and convertible notes. Besides, it requires lawyers to manually verify spreadsheets, which may lead to more human errors and difficulty in leaving audit trails. With Nasdaq Linq, private stock issuers own digital ownership so that the settlement time can be greatly reduced. Chain pointed out that the current standard settlement time for the equity trading market is three days, whereas the application of blockchain technology can decrease the settlement time to 10 minutes, as well as reduce security risk by 99$\%$ \cite{ref68}, significantly improving the efficiency. Online completion of issuance and subscription by both parties can also effectively simplify unnecessary paperwork, and reduce the administrative risk and burden faced by issuers due to the heavy approval process. The blockchain-based private equity trading platform of Nasdaq Linq provides companies with a dashboard to manage valuations, an equity change timeline chart, and investor personal equity certificates, enabling issuers and investors to better track and manage relevant information. The use of blockchain technology replaces the traditional methods of paper and electronic spreadsheets, greatly improving transaction and management efficiency \cite{ref69}.

\begin{figure}
	\includegraphics[scale=0.7]{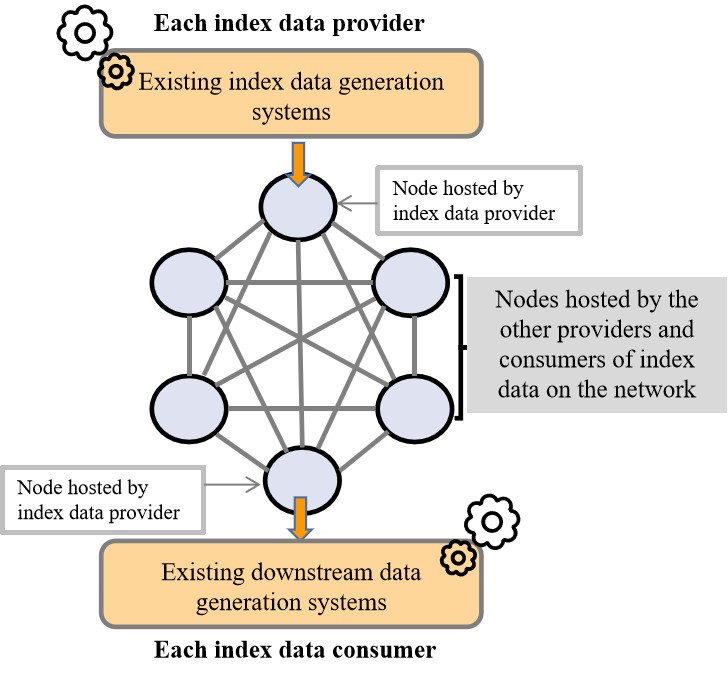}
	\centering
	\caption{The inner workings of the Indexed Data Network.}
	\label{fig10}
\end{figure}
\subsubsection{Clearing and Settlement System in the Securities Industry}
\ 
\newline 
\newline The clearing and settlement system is a system used for securities transactions settlement, which involves two main processes: clearing and settlement. Clearing is the process of matching, verifying, and reconciling trade details between the buyer and the seller. Settlement is the process of actually transferring the agreed-upon financial assets (e.g., stocks, bonds, cash) from the seller to the buyer and vice versa. Traditional securities settlement systems rely on centralized securities exchanges and securities clearing organizations to complete securities clearing and settlement. During the trading process, buyers and sellers of securities need to submit trading orders to the securities exchange through brokers, and the settlement and delivery of securities also need to be completed through the securities clearing organization. The entire process requires the participation of multiple intermediaries, which prolongs settlement time and introduces risks associated with multiple intermediaries. However, blockchain technology can bring many benefits to the securities clearing and settlement system, such as improving trading efficiency, reducing trading costs, and enhancing trading security.

Goldman Sachs is one of the world's most powerful investment banks, and it is also the earliest financial institution to research blockchain technology. It proposed an application of "Securities Settlement Cryptocurrency" (known as  SETLcoin), which is designed to utilize blockchain technology for securities trading and settlement, enabling users to trade financial assets using SETLcoin via a virtual wallet. The virtual wallet for SETLcoin is built on blockchain technology, and each user can have a virtual account to store and manage their financial assets including stocks, bonds, currencies, and more. Whenever a user attempts to make a trade, SETLcoin will exploit smart contracts to validate the transaction's legality and record transaction data. Subsequently, this transaction data will be stored in a distributed ledger accessible and verifiable by each node in the blockchain network \cite{ref71}\cite{ref70}.

It’s worth noting that SETLcoin is developed as a clearing and settlement system based on the Bitcoin blockchain, rather than a true cryptocurrency. It is just a token used to represent specific securities and does not have the characteristics of Bitcoin and other Cryptocurrencies as payment means. SETLcoin employs the PBFT (Practical Byzantine Fault Tolerance) consensus algorithm to ensure consistency among all nodes. In PBFT, nodes vote to determine which transaction records could be added to the blockchain. Transaction records can only be added to the blockchain when the majority of nodes agree. Furthermore, SETLcoin adopts sidechain technology to improve transaction speed and scalability. A sidechain is a blockchain that runs parallel to the main blockchain, but it can process more transaction records. SETLcoin utilizes sidechains to handle a large number of small transactions while putting large transaction records on the main blockchain for processing. This design effectively improves transaction processing speed and throughput. SETLcoin also employs multi-signature technology to enhance security. In multi-signature technology, transactions require multiple signatures to be confirmed and added to the blockchain, preventing unauthorized personnel from tampering with or maliciously attacking transactions. In conclusion, using SETLcoin, securities transactions can be settled in seconds instead of waiting for the traditional settlement cycle of days.


\subsection{Financial Analysis}
Financial analysis in the securities industry refers to the in-depth analysis of financial markets and securities trading, aiming to provide decision support and advice for investors and financial institutions. Traditional financial analysis bears their security risks and challenges in certain aspects, such as data security and credibility issues, transaction transparency issues, and data analysis efficiency issues. Blockchain technology can provide better solutions to these problems with two key properties: immutability and tamper-resistance. 

The properties of blockchain technology have made it the one of most popular technologies in the field of financial analysis. In particular, the securities industry has widely adopted blockchain technology for data management, improving the processing of securities analysis. Symbiont is a leading blockchain-based financial company that utilizes a blockchain platform for securities analysis. This platform first utilizes machine learning and artificial intelligence technologies to automatically identify and analyze valuable information from huge data. Then it offers asset management functions for securities issuers via blockchain, which has improved transparency, thus achieving better market information and data analysis tools for investors.

Symbiont provides a blockchain network that supports the connection of data providers and consumers. Specifically, the decentralized network powered by Symbiont Assembly™ \cite{ref103} allows parties to share accurate and auditable data in real time. Symbiont Assembly™ powers a live index data network that expedites data delivery, which eliminates the need for manual updates and thus reduces risks. The overview of its Index Data network framework is illustrated in Fig.\ref{fig10}, where index data providers transfer data to nodes hosted by index providers through an existing index data generation system. After submitting the data, automatic data verification checks are performed using smart contract applications. The data is then encrypted, written to a log on the node hosted by the index data provider, and shared in real time with all nodes on the network. Only nodes managed by permissioned data consumers will have the decryption keys needed to access a given dataset. After decrypting the index data, data consumers pass it to downstream systems.

To sum up, the blockchain-based financial analysis platform provides a decentralized, non-tamperable, credible solution for data management. Since the transaction records recorded on the blockchain are traceable, the transaction process is more transparent and regulated. In addition, smart contract brings data analysis more benefits including accuracy, efficiency, trust, and security.

\begin{figure*}[htb!]
	\includegraphics[scale=0.55]{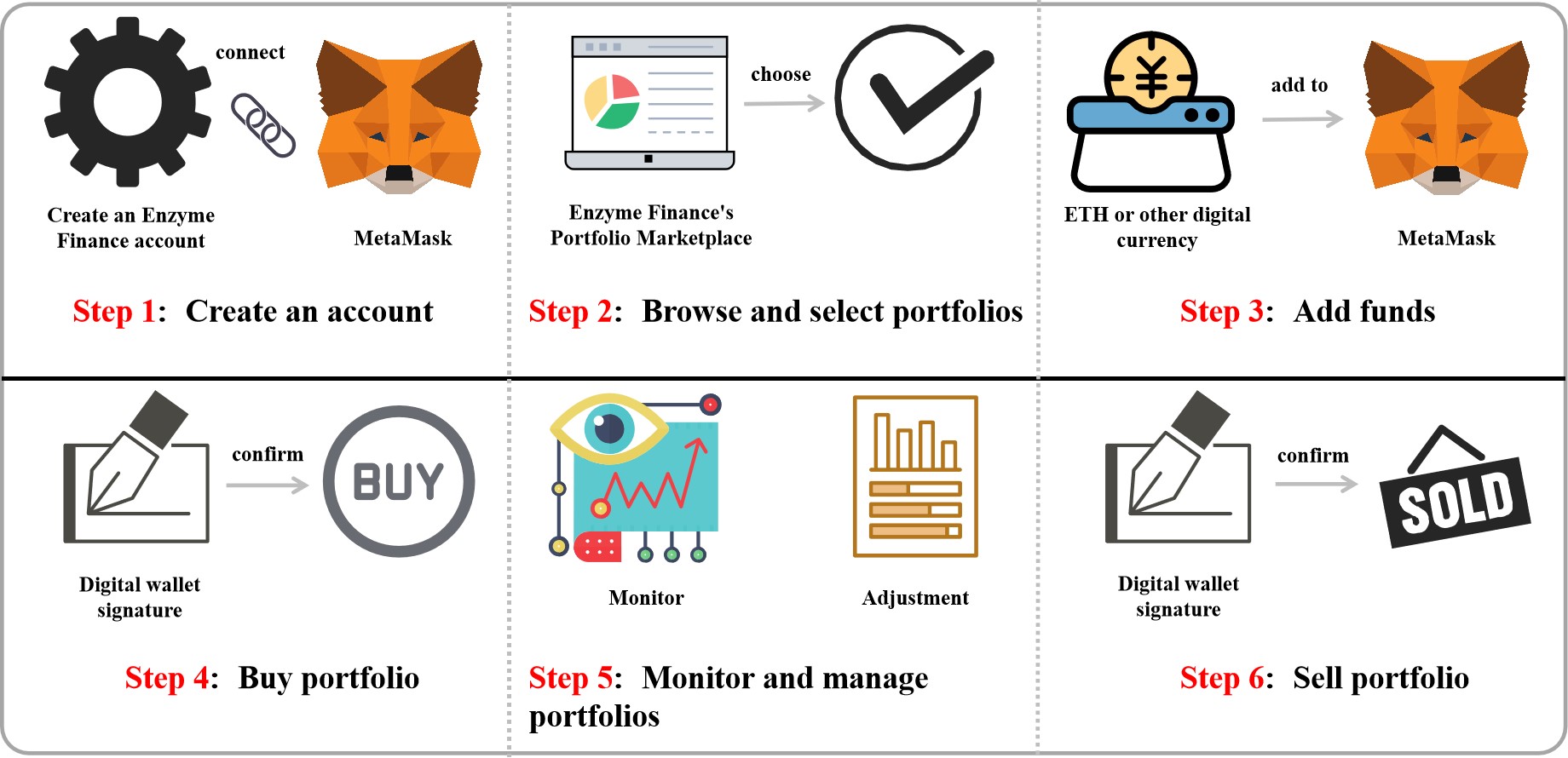}
	\centering
	\caption{Enzyme Finance digital asset investment process.}
	\label{fig11}
\end{figure*}

\subsection{Investment Management}

Investment management plays a crucial role in the securities industry. It refers to the behavior of investment managers entrusted by investors, including investment management services of securities or other financial products for specific goals and benefits of investors, which is paid by management fees. Investment managers require specialized knowledge and skills in portfolio construction, asset allocation, risk management, market analysis, and client relationship management. As the investment is made by entrusted managers, the process of investment may incur many problems such as high management costs, opaque investment management services, and inflexible investment portfolios.

With the rapid development of blockchain technology, more and more financial institutions are exploring the advantages of blockchain technology in the securities industry. Enzyme Finance \cite{ref102}, a digital asset management platform, utilizes Ethereum blockchain technology and smart contracts to provide efficient, secure, and decentralized digital asset management services.

Fig.\ref{fig11} shows the process of investing in digital assets provided by Enzyme Finance. First, users need to create an Enzyme Finance account and connect their digital wallet (such as MetaMask or Ledger) to trade and manage digital assets on the platform( \emph{i.e.}, Step 1 in Fig.\ref{fig11}). Secondly, users browse the Enzyme Finance portfolio market to view portfolios created by other investors and learn about their asset types, weight allocations, and historical performance. By using various functions (like searching and filtering) on the platform, users can find portfolios of interest(\emph{i.e.}, Step 2 in Fig.\ref{fig11}). Subsequently, users need to deposit ETH or other supported digital currencies into their digital wallet for the purpose of purchasing the selected portfolio(\emph{i.e.}, Step 3 in Fig.\ref{fig11}). Then, users use the purchase function to confirm the purchase quantity and price and sign the transaction with their digital wallet to complete the purchase(\emph{i.e.}, Step 4 in Fig.\ref{fig11}). Once the purchase is complete, users employ the portfolio management function to monitor the performance, configuration, and risk control of their portfolio. In addition, users are allowed to add or withdraw funds at any time and adjust both the weight allocation and strategy of their portfolio as needed(\emph{i.e.}, Step 5 in Fig.\ref{fig11}). Finally, when users determine to dump their portfolio, the sell function offered by the platform could be selected to change their held digital assets. In particular, users can confirm the quantity and price of the sale and sign the transaction with their digital wallet to complete the sell-off (\emph{i.e.}, Step 6 in Fig.\ref{fig11}).

To summarize, Enzyme Finance is an innovative digital asset management platform that utilizes blockchain technology to provide efficient, secure, and decentralized digital asset management services. The digital asset management services provided by Enzyme Finance are decentralized, thus investors can directly control their digital assets without the help of traditional financial institutions. By using smart contracts to specify portfolio construction and management, investors can achieve more flexible asset allocation and management on the Enzyme Finance platform. As such, Enzyme Finance offers better investment management experience for both investment managers and investors.

\begin{table*}[htb!]
	\caption{ \textbf{Comparison of the main characteristics of six blockchain implementations.}}
	\label{C}
	\centering
	\resizebox{2.0\columnwidth}{!}{
	\begin{tabular}{lllllll}
		\hline
		\begin{tabular}[c]{@{}l@{}}Characteristics\\ /Platforms\end{tabular} & Ethereum                                                                               & Hyperledger Fabric                                                                & Corda                                                                   & Quorum                                                                                 & Symbiont Assembly™                                                                           & tZERO                                                           \\ \hline
		\begin{tabular}[c]{@{}l@{}}Platform \\ Description\end{tabular}      & \begin{tabular}[c]{@{}l@{}}A public \\ blockchain \\ platform\end{tabular}             & \begin{tabular}[c]{@{}l@{}}Business-to-Business\\ centric blockchain\end{tabular} & \begin{tabular}[c]{@{}l@{}}Financial-focused\\ DLT\end{tabular}         & \begin{tabular}[c]{@{}l@{}}Financial-focused\\ DLT (built on \\ Ethereum)\end{tabular} & \begin{tabular}[c]{@{}l@{}}Financial-focused\\ DLT\end{tabular}                              & \begin{tabular}[c]{@{}l@{}}Financial \\ -focused\\ DLT\end{tabular} \\
		Governance                                                           & \begin{tabular}[c]{@{}l@{}}Ethereum \\ developers\end{tabular}                         & Linux Foundation                                                                  & R3                                                                      & ConsenSys                                                                              & Symbiont                                                                                     &     \begin{tabular}[c]{@{}l@{}}tZERO Group,\\ Inc\end{tabular}                                            \\
		\begin{tabular}[c]{@{}l@{}}Blockchain \\ Type\end{tabular}           & Public                                                                                 & Private                                                                           & Private                                                                 & Private                                                                                & Private                                                                                      & Private                                                         \\
		Access Type                                                          & Permissionless                                                                         & Permissioned                                                                      & Permissioned                                                            & Permissioned                                                                           & Permissioned                                                                                 & Permissioned                                                    \\
		\begin{tabular}[c]{@{}l@{}}Consensus \\ Mechanism\end{tabular}       & PoW                                                                                    & Multiple                                                                          & \begin{tabular}[c]{@{}l@{}}Implementations\\ (NotaryNodes)\end{tabular} & RAFTIBFT, PoA                                                                          & \begin{tabular}[c]{@{}l@{}}state machine \\ replication\end{tabular}                         & Tzero ATS                                                       \\
		Smart Contract                                                       & Yes                                                                                    & Yes                                                                               & Yes                                                                     & Yes                                                                                    & Yes                                                                                          & Yes                                                             \\
		\begin{tabular}[c]{@{}l@{}}Digital \\ Currency\end{tabular}          & \begin{tabular}[c]{@{}l@{}}Ethers and \\ tokens through\\ smart contracts\end{tabular} & \begin{tabular}[c]{@{}l@{}}No native asset,\\ Internal token\end{tabular}         & Native token, XDC                                                       & None                                                                                   & \begin{tabular}[c]{@{}l@{}}No native asset,\\ support Bitcoin,\\ Ethereum, etc.\end{tabular} & TZROP      \\ 
		\begin{tabular}[c]{@{}l@{}}Proportion of \\Use by Top\\ 100 Companies\cite{ref177}\end{tabular}          & 24\% &  38\%        &        13\%                                             & 17\%                                                      & NONE & NONE \\ \hline                                                    
	\end{tabular}  }
\end{table*}

\begin{table*}[htb!]
	\caption{List of events of mainstream financial institutions in the blockchain field.}
	\label{D}
	\centering
	\resizebox{1.7\columnwidth}{!}{
		\begin{tabular}{|p{3cm}|p{3cm}|p{10.5cm}|}
			\hline
			\multicolumn{1}{|c|}{Organization Type}             & \multicolumn{1}{c|}{Organization}               & \multicolumn{1}{c|}{Event}  \\ \hline
			\multirow{4}{*}[-6ex]{Central Bank}                                                                    & The People's Bank Of China    & In 2017, the People's Bank of China launched collaborative research and development initiatives aimed at the e-CNY, engaging select commercial banks and relevant institutions.                                                                                                                                                                                                                                                                           \\ \hhline{~--} 
			&  Russia          & The Central Bank of Russia established a task force to explore domains covering distributed ledger technology and payment methods \cite{ref152}.
			\\ \hhline{~--} 
			& European Central Bank                                                                & The European Central Bank focused on the application of blockchain technology in securities and payment settlement systems \cite{ref153}.                                                                                                                           \\ \hhline{~--} 
			& Bank Of England                                                                      & The Bank of England collaborated with academic institutions to develop the digital currency(RSCoin), which has currently progressed to the testing phase \cite{ref154}.                                                                                                                                                                                                                                               \\ \hline
			\multirow{2}{*}[-4ex]{\begin{tabular}[c]{@{}c@{}}Securities Market \\ Intermediaries\end{tabular} }                                              & Deloitte                                                                             &  Deloitte has maintained a consistent commitment to exploring solutions within the audit industry that are built upon blockchain technology. Previously, Deloitte conducted research on more than 20 distinct use cases for blockchain technology \cite{ref155}.                     \\ \hhline{~--} 
			& Goldman Sachs                                                                        &  Based on the Bitcoin blockchain, Goldman Sachs  developed a system for settlement of securities transactions through cryptocurrencies, called SETLcoin \cite{ref156}.                                                                                                                                                                                                                \\ \hline
			
			\multirow{5}{*}[-20ex]{Commercial Bank}                                                                 & Bank Of America                                                                      & In the domain of blockchain technology, Bank of America has submitted patent applications totaling more than 40 and has also played a substantial role in numerous alliances and organizations closely linked to the blockchain field \cite{ref157}.                                                                                                                                                              \\ \hhline{~--} 
			& UBS                                                                                  & UBS Group has inaugurated a new technology research center in London, with a dedicated focus on exploring the integration of blockchain technology into financial operations \cite{ref158}. \\ \hhline{~--} 
			& ANZ Bank                                                                             &  ANZ Bank has integrated blockchain technology into its internal operations to streamlineprocesses, mitigate risks and enhance the experiences of both employees and customers. By harnessing Distributed Ledger Technology (DLT), they have successfully implemented automated processes for Bank Trade Risk Participation Sales (RSPD), leading to a notable reduction in operational risks and relieving the bank's product portfolio management and operational teams of certain burdens \cite{ref159}.                                                                                                                                                                                                                                                                                                                                                                             \\ \hhline{~--} 
			&  Commonwealth Bank of Australia             & The Commonwealth Bank of Australia (CBA) has been actively delving into the practical applications of blockchain technology. Notably, the bank has been consistently involved in the exploration of various blockchain use cases for over four years, culminating in the successful completion of 25 concept validations and tests \cite{ref160}.                                                                                                                                                                                                                                                                                                                                                 \\ \hhline{~--} 
			& Spanish Santander Bank                                                               &  In April 2018, Banco Santander, based in Spain, introduced the blockchain-powered cross-border forex trading application named "One Pay FX." This innovative application enables real-time cross-border remittances for users spanning Spain, the United Kingdom, Brazil, and Poland \cite{ref161}.                                                                \\ \hline
			\multirow{2}{*}[-4ex]{\begin{tabular}[c]{@{}c@{}}Securities Issue \\ Regulatory Authority\end{tabular}} & SEC                                                                                             &The U.S. Securities and Exchange Commission has approved the online retailer Overstock.com's issuance of its new publicly listed stocks on the blockchain \cite{ref162}.                                                                                                                                                                                                                                                                                                                                \\ \hhline{~--} 
			&   European Securities and Markets Authority   &  In 2020, the European Commission presented a draft of the Markets in Crypto-Assets Regulation (MiCA), with the goal of addressing market volatility risks, money laundering, and terrorism financing issues arising from crypto-assets and decentralized finance (DeFi) activities \cite{ref163}.                                                                                        \\ \hline   
			
			\multirow{10}{*}[-20ex]{\begin{tabular}[c]{@{}c@{}}Stock Market \\ Infrastructure\end{tabular}}                                                     & Depository Trust \& Clearing  Corporation            & The Depository Trust \& Clearing Corporation (DTCC) issued a white paper entitled "Tapping the potential of distributed ledgers to improve the post-trade landscape" This document provides a comprehensive exposition of DTCC's viewpoint regarding the implementation of blockchain technology within the securities trading industry \cite{ref164}.                                                                                                                                                                                                                                                                                                                                                                                                                                                                                                                                            \\ \hhline{~--} 
			& Nasdaq                                                                                          & Nasdaq introduced Nasdaq Linq, a private equity trading platform developed in collaboration with the blockchain startup Chain.com \cite{ref165}.                                                                                                                                                                                                                                                                                                                                                                                                                                                                                                      \\ \hhline{~--} 
			& Australian Securities Exchange                       &  The Australian Securities Exchange (ASX)employs the blockchain ledger technology developed by Digital Asset to replace the current CHESS post-trade settlement system, while concurrently managing the clearing and settlement of stock transactions \cite{ref166}.                                                                                                                                                                                                                                                                                                                                                                                                                                                                                                           \\ \hhline{~--} 
			& Deutsche Börse                                                                                  & In partnership with the German Central Bank, Deutsche Börse created a blockchain prototype using the Hyperledger project, successfully incorporating functionalities like electronic securities, digital currency settlement, and bond repurchases \cite{ref167}. In 2018, Deutsche Börse partnered with HQLAx to engage in the research and development of a blockchain-based securities lending solution, with the objective of bolstering securities collateral liquidity on a worldwide scale \cite{ref168}.                                                                                 \\ \hhline{~--} 
			& London Stock Exchange                                                                           &  The London Stock Exchange (LSE) is actively driving exploration in the issuance and trading of digital securities and assets. The LSE made an investment in the blockchain company Nivaura and collaborated to develop a decentralized platform tailored for the issuance of security tokens \cite{ref169}.                                                                           \\ \hhline{~--} 
			& Singapore Exchange                                                                              &  In 2016, the Singapore Exchange became involved in the financial technology project Ubin. By July 2020, all five project phases had been successfully completed, effectively confirming the viability of employing blockchain for cross-border settlement and payment processing \cite{ref170}.               \\ \hhline{~--}
			& Hong Kong Stock Exchange   & The Hong Kong Stock Exchange and the Australian Stock Exchange collaborated to leverage their experience in settling transactions within blockchain systems, with the aim of applying blockchain technology to stock lending and over-the-counter trading operations \cite{ref171}. \\ \hhline{~--} 
			
			& Australian Stock Exchange  & The Australian Stock Exchange enlisted blockchain startups as part of its initiative to build a settlement system founded on distributed ledger technology \cite{ref172}.                                                                                                                                                                                                                                                                                                                                                                                                                                                                                                                                                                                                                                                                                                                \\ \hhline{~--} 
			& New York Stock Exchange       & In 2015, the New York Stock Exchange made an investment in the cryptocurrency exchange Coinbase and later launched the world's inaugural Bitcoin index NYXBT, which was issued by a securities exchange \cite{ref173}.                                                                                                                                                                                                                                                                                                                                                                                                                                                                                                             \\ \hhline{~--} 
			& Intercontinental Exchange     &  Intercontinental Exchange partnered with blockchain startup enterprises to introduce the real-time Cryptocurrency Data Feed service, designed for monitoring market data associated with digital currencies. They successfully launched the global digital asset trading platform named Bakkt \cite{ref174}. 
			\\    \hline                
	\end{tabular}}
\end{table*}

\section{Observations}
To better analyze the research status of blockchain applications in finance areas, we observe from the following three perspectives: (1) comparative analysis of blockchain platforms, (2) research events of mainstream financial institutions, and (3) security issues of Decentralized Finance Applications (DeFi).
\subsection{Comparative Analysis of Blockchain Platforms}

In this section, we will compare the blockchain platforms introduced in the previous chapter, and examine their strengths and weaknesses. This is done to summarize what adjustments are required to apply blockchain in financial areas. Table \ref{C} provides a summary comparison of key features of six blockchain implementations.

Ethereum is a decentralized, public blockchain platform that supports smart contracts and is renowned for its extensive developer community and wide range of applications. It is a leader in various fields such as DeFi and NFT. However, the current transaction speed of Ethereum cannot fully meet the needs of the financial field, and the high fees of Ethereum make it unacceptable for ordinary people. At the same time, Ethereum still has room for improvement in terms of security and privacy.
Hyperledger Fabric is a blockchain platform for industries such as finance, supply chain, and healthcare; Corda is mainly designed for the financial service industry; Quorum is based on Ethereum and mainly serves the financial industry; Symbiont Assembly serves the capital market; Tzero Focus on the trading of security tokens and non-fungible tokens (NFT). 

Upon observation, it can be noted that apart from Ethereum, all other platforms are permissioned blockchains. As discussed by \cite{ref126}, public blockchains are highly suitable for creating global and uncensored payment solutions, such as Bitcoin and Monero. In contrast, permissioned blockchains are better suited for applications involving smart contracts, and they are also more suitable for enterprise use, particularly in the financial sector. In addition, they have faster transaction speed, better scalability, higher security, and privacy protection.

Permissioned blockchain systems have higher security compared to public blockchain systems since they require an access control layer. Only authorized participants are allowed to join the network and access data. This is especially important in the financial industry where financial data contain sensitive information and financial transactions involve a lot of money. Additionally, permissioned blockchain systems do not use the proof-of-work mechanism. As a result, these blockchain systems can improve trading efficiency in the financial sector by enabling faster transaction confirmations, lower transaction costs, and simplified audit as well as compliance processes. Finally, permissioned blockchain systems can be customized and optimized according to the detailed needs of varying financial scenarios, such as selecting appropriate consensus mechanisms, privacy protection mechanisms, smart contract languages, etc.

\subsection{Research Events of Mainstream Financial Institutions}

In recent years, a great number of innovative institutions have begun exploring blockchain technology in the financial industry. At the same time, a significant amount of financial institutions are actively investing in blockchain technology enterprises. Table \ref{D} provides an overview of research events of mainstream financial institutions.

We divide all research events into three types according to different attitudes towards blockchain of these institutions: \textit{continue to pay attention, participate in research,} and \textit{invest in blockchain startups}. The specific description is as follows:

\begin{itemize}
		\item[	1).]\textbf{Continue to pay attention:} Regulatory agencies of major economies are highly concerned with the development and application of blockchain technology in the financial sector. They have recognized both its enormous potential and prospects, as well as its security risks and regulatory challenges. Several institutions consider that the current blockchain systems are not mature and only by strengthening regulation can the healthy development of blockchain applications be promoted.
		\item[	2).]\textbf{Participate in research:} Financial institutions have actively attempted to explore blockchain technology in their own business areas. Prominent institutions such as Bank of America and Goldman Sachs have already begun actively reserving blockchain-based technology patents. These institutions believe that blockchain technology has the potential to transform the existing financial system, improve the efficiency and security of financial transactions bring new business opportunities to the financial industry. Therefore, they have started to explore and invest in the blockchain field, hoping to gain a first-mover advantage in future competition.
		\item[	3).]\textbf{Invest in blockchain startups:} In addition to actively exploring the application of blockchain technology using their own resources, financial institutions are also actively investing in or partnering with these startups in various ways. Financial institutions believe that by working with these startups, they can more quickly apply blockchain technology to their own businesses and share the accumulated technology and experience of these startups. For example, well-known financial institutions such as UBS Group, Citigroup, and JPMorgan Chase have invested in blockchain technology startups. These institutions believe that investing in these startups can bring more technological and market opportunities, and maintain a competitive advantage in the blockchain technology industry.
		
  \end{itemize}

\subsection{Security Issues of Decentralized Finance Applications (DeFi)}

\begin{figure*}[htb!]
	\includegraphics[scale=0.5]{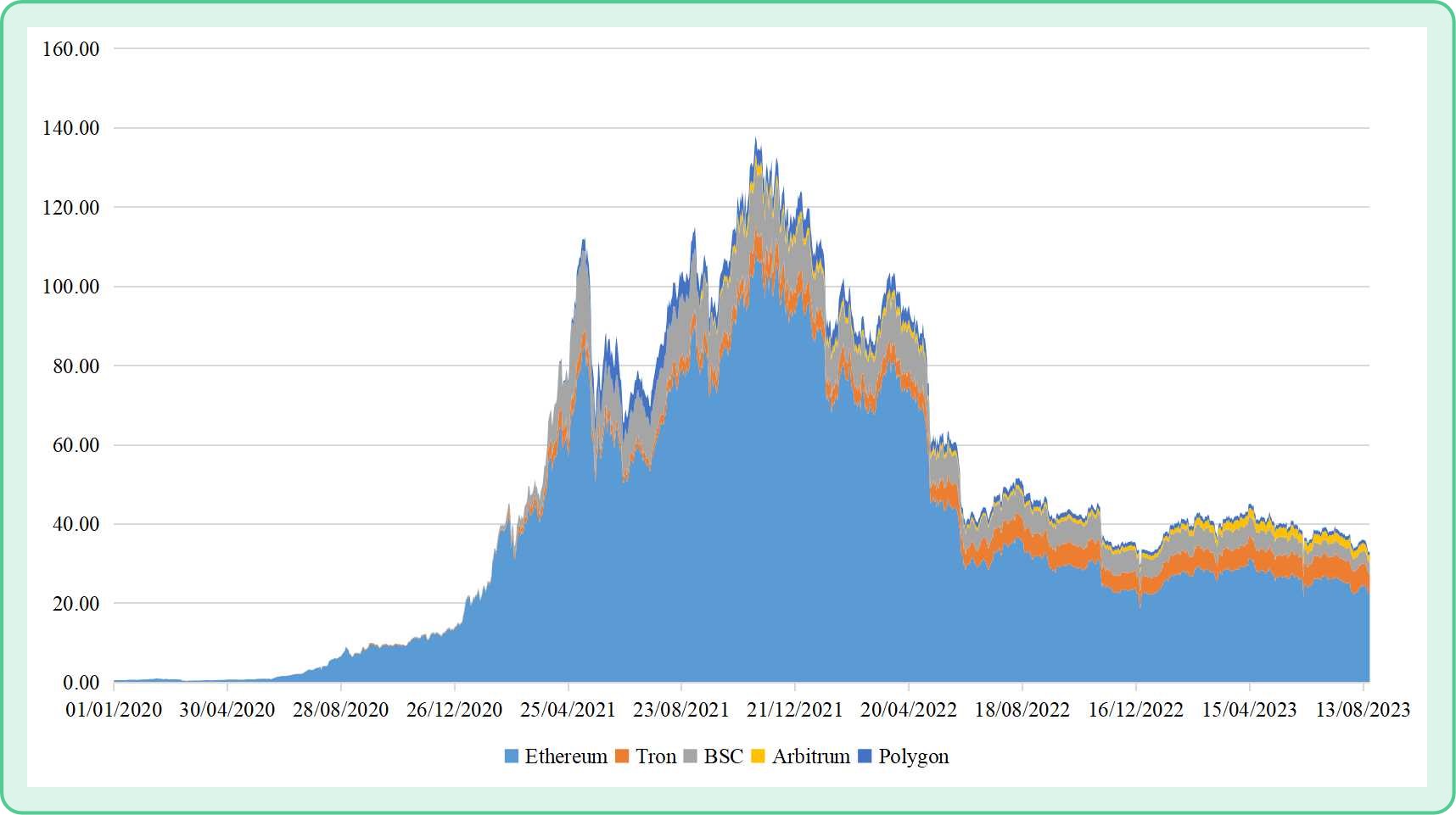}
	\centering
	\caption{The total value (in billions of US dollars) of all assets locked in DeFi contracts on major blockchain platforms from January 2020 to August 2023.}
	\label{5}
\end{figure*}

Decentralized finance applications (DeFi) represent an innovative form of financial application that leverages decentralized blockchain technology \cite{ref72}\cite{ref73}. Recently, DeFi has emerged as one of the most popular application types on public blockchains. Decentralized finance applications are typically composed of numerous smart contracts, enabling the creation of versatile financial services that operate on the blockchain \cite{ref74}\cite{ref75}. Fig.\ref{5} depicts the cumulative value (in billions of US dollars) of all assets locked in DeFi contracts on major blockchain platforms from January 2020 to August 2023. As the range of use cases for decentralized finance continues to expand, the total value of assets locked in DeFi has witnessed a substantial growth from \$675 million in January 2020 to \$122 billion in February 2023 \cite{ref81}.

Presently, numerous financial services have transitioned into the DeFi ecosystem. Through a comprehensive summary and analysis of existing works \cite{ref131}\cite{ref132}\cite{ref133}\cite{ref134}\cite{ref135}\cite{ref136}\cite{ref137}\cite{ref139}\cite{ref141}\cite{ref143}, we categorize the applications of DeFi into six categories.

\begin{itemize}
	\item[	1)]\textbf{Lending and Borrowing:} Assets in a DeFi application are lent and borrowed using protocols designed for fund loans, commonly known as DeFi lending protocols \cite{ref134}\cite{ref143}. Among DeFi applications, decentralized lending services form the most substantial category, boasting a total value locked exceeding \$40 billion. These services extend loans to individuals or businesses utilizing smart contracts as automated agents or intermediaries, streamlining the lending and borrowing processes.
	\item[	2)]\textbf{Decentralized Exchange:} Decentralized Exchange (DEX) is fundamentally a type of DeFi project facilitating on-chain digital asset exchanges \cite{ref135}\cite{ref143}. Users engage in decentralized trading of various tokens through interaction with smart contracts. The cumulative locked funds in decentralized exchanges have surpassed \$25 billion. For example, \textbf{Uniswap}, whose users lock up tokens worth approximately \$8 billion, is one of the biggest DEXs. What sets Uniswap apart is its innovative adoption of an Automated Market Maker (AMM) design \cite{ref130}. This design eliminates the need for traditional order books and relies instead on smart contracts and liquidity pools to facilitate transactions \cite{ref128}\cite{ref129}. This means that anyone can become a liquidity provider by depositing funds into these liquidity pools, thereby supplying the necessary liquidity for trades.
	\item[	3)]\textbf{Portfolio Management:} With an increasing number of DeFi projects encouraging clients to contribute liquidity, a novel category of projects, referred to as portfolio management, has emerged to assist users (\emph{i.e.}, liquidity providers) in investing their assets \cite{ref136}\cite{ref143}. These projects autonomously identify DeFi projects offering the highest annual percentage yield (APY).  
	\item[	4)]\textbf{Derivative:} DeFi derivatives are created using smart contracts that derive their value from the performance of an underlying entity, such as currencies, bonds, and interest rates \cite{ref137}\cite{ref143}. Tokenized derivatives can be generated without the need for trusted third parties, thereby mitigating the potential influence of malicious attacks. Despite approximately 99\% of derivative trading volume occurring on centralized exchanges, a growing number of DeFi projects have surfaced, offering comparable functionality, particularly in futures, perpetual swaps, and options \cite{ref138}.
	\item[	5)]\textbf{Stablecoin:} Stablecoins represent a category of cryptocurrencies engineered to ensure price stability \cite{ref139}\cite{ref143}. Typically, these coins achieve stability through direct/indirect backing or intervention via various stabilization mechanisms. Well-known stablecoins like USDC or USDT are custodial and fall outside the realm of DeFi, as they predominantly depend on a trusted third party. In decentralized environments, the challenge for protocol designers lies in creating a stablecoin that attains price stability in an economically secure and consistent manner, allowing all necessary parties to participate profitably \cite{ref140}. Price stability is pursued through on-chain collateral, forming the basis for secured loans that underpin the stablecoin's economic value. Non-custodial stablecoins aim to operate independently of the societal institutions on which custodial designs rely.
	\item[	6)]\textbf{Aggregator:} A DeFi aggregator serves as a platform that consolidates trades from various decentralized platforms into a single interface, enhancing efficiency for cryptocurrency transactions \cite{ref141}\cite{ref143}. Typically, a DeFi aggregator utilizes multiple decentralized exchanges (DEXs) and deploys diverse buying and selling strategies to assist users in maximizing profits while mitigating gas fees and DEX trading fees. These aggregators not only aggregate the best prices but also provide a unique, user-friendly approach for analyzing and combining other users' trading strategies through a convenient drag-and-drop mechanism \cite{ref142}. With the introduction of DeFi aggregators, newcomers to the industry can leverage DeFi benefits without delving into the intricacies of trading technologies, decentralized services, blockchain, \emph{etc}. Overall, an aggregator contributes to users making more informed trading decisions.
\end{itemize}



	The decentralized nature of DeFi offers numerous benefits while entailing risks. DeFi, holding trillions of dollars, has become an appealing target for numerous external attackers, posing serious threats to its applications. In 2021 alone, DeFi users incurred losses exceeding 10 billion USD due to the attacks. Table \ref{E} presents well-known DeFi projects that have fallen victim to such attacks. We have summarized four key factors that could cause these attacks: (1) Coding errors or logical loopholes within smart contracts create windows of opportunity for attackers to exploit vulnerabilities. (2) The abuse of third-party protocols allows attackers to manipulate transactions by leveraging external data or interfaces. (3) Improper use of flash loans grants attackers the ability to manipulate markets or swiftly seize rewards. (4) The tampering or hijacking of frontend interfaces empowers attackers to deceive users by prompting them to input incorrect information or redirecting them to malicious websites.

 Obviously, despite the popularity of DeFi, DeFi is still in its infancy phase. In order to tackle DeFi attacks, we propose three strategies: (1) DeFi project developers should conduct comprehensive testing and auditing prior to the release of DeFi. This ensures that the code is devoid of vulnerabilities or defects, adheres to established practices and standards, and utilizes trustworthy third-party protocols or services. (2) DeFi project teams and developers should actively monitor and analyze the dynamics of the DeFi market and network subsequent to the release and operation of DeFi protocols or applications. This includes promptly identifying and reporting any unconventional or suspicious transactions, as well as employing professional tools and platforms to prevent potential attacks. (3) DeFi project teams and developers should enact immediate emergency measures, such as suspending or upgrading protocols or applications, notifying users and the broader community, tracking and penalizing attackers, and compensating users for any incurred losses.
	
	\begin{table}[htb!]
		\caption{\textbf{The statistics on well-known Defi projects that 
  have been attacked.}}
		\label{E}
		\centering
		\begin{tabular}{l|l|l}
			\hline
			\hline
			\makebox[0.15\textwidth][l]{\textbf{Project Name} }            & \makebox[0.15\textwidth][l]{\textbf{Attacking Time}}          & \makebox[0.1\textwidth][l]{\textbf{Loss}} \\ \hline
			Balancer               & 2020.06.28  & \$0.5 Million    \\ \hline
			MakerDAO               & 2020.03.12    & \$9 Million    \\ \hline
			ChainSwap              & 2021.07.11  &  \$4.8 Million   \\ \hline
			Burgerswap             & 2021.05.28  &  \$7.2 Million   \\ \hline
			bZx-V2                 & 2020.09.15  &  \$8.1 Million   \\ \hline
			bZx-V1                 & 2020.02.17  &   \$9.4 Million  \\ \hline
			Akropolis              & 2020.11.12    &  \$2 Million   \\ \hline
			EasyFi                 & 2021.04.20   &  \$80 Million   \\ \hline
			Eminence Finance       & 2020.09.29   &   \$15 Million  \\ \hline
			JulSwap                & 2021.05.28  & \$0.7 Million    \\ \hline
			Furucombo              & 2020.09.29  &   \$15 Million  \\ \hline
			Harvest Finance             & 2020.10.26 &   \$33.8 Million  \\ \hline
			Grim Finance            & 2021.12.19 &   \$30 Million  \\ \hline
			Indexed Finance            & 2021.10.14 & \$16 Million    \\ \hline
			Lendf.Me            & 2020.04.18 &  \$25 Million   \\ \hline
			Nerve                & 2021.11.15 &  \$8 Million   \\ \hline
			Origin                 & 2020.11.17 &   \$7 Million  \\ \hline
			Oypn                   & 2020.08.04 &   \$0.37 Million  \\ \hline
			PancakeBunny          & 2021.05.20 &  \$47.1 Million   \\ \hline
			PAID Network           & 2021.03.05 &   \$160 Million  \\ \hline
			Pickle Finance           & 2020.11.22 &   \$20 Million  \\ \hline
			Poly Network         & 2021.08.10 &  \$611 Million   \\ \hline
			Popsicle         & 2021.08.04 &   \$25 Million  \\ \hline
			Rari Capital         & 2022.04.30 &   \$90 Million  \\ \hline \hline
		\end{tabular}
	\end{table}

\section{Conclusion}\label{sec10}

Currently, blockchain technology has infused new vitality into financial transactions. The digitized and decentralized securities industry will bring about another financial infrastructure revolution. This paper provides a comprehensive review of the principles of blockchain technology and its applications in the financial sector. Particularly, we spare more efforts on exploring blockchain applications in the securities industry.

To begin with, upon comparing different blockchain platforms used in the financial application field, we have identified shortcomings in the business flexibility and decision-making efficiency of permissionless blockchains. Conversely, permissioned blockchains, which are controlled by specific institutions, can improve the operational efficiency of existing financial institutions and hold greater practical significance within the current legal and business environment.

Secondly, we systematically outline the four main directions in which blockchain technology is being applied in finance areas: \textit{capital raising, securities trading, financial analysis,} and \text{investment management}. The inherent characteristics of \textit{immutability, transparency,} and \textit{high security} make blockchain naturally suitable for the securities industry. Blockchain technology can achieve real-time recording and sharing of trading data, ensuring the traceability and transparency of transaction information, and effectively preventing fraud and tampering. Furthermore, the smart contract automates the execution of transactions and settlements. As a result, costs and risks associated with intermediate links are reduced, ultimately enhancing transaction efficiency. Therefore, blockchain technology holds vast potential for application in the securities industry, promising to provide robust support for the stable development of the securities market.

Lastly, we review the current state of blockchain applications in the financial sector from the perspective of DeFi. We observe that DeFi, as a revolutionary idea, has grown incredibly popular in recent years and offers an alternative financial ecosystem that subverts centralized systems. We also find that there are still some shortcomings of DeFi in both business processes and technical aspects. Blockchain technology is still in its early stages. Consequently, it is crucial to continue promoting innovative techniques and ideas in blockchain while simultaneously reinforcing regulation and standards to ensure safety in the financial industry. Only in this way can we realize the true value of blockchain technology in the financial sector.

\section{Acknowledgements}

This work is supported by the National Key R\&D Program of China (No. 2021YFB2700500), the Key R\&D Program of Zhejiang Province (No. 2023C01217), and the National Natural Science Foundation of China (No. 62372402).



\bibliographystyle{plain}
\bibliography{references.bib}

\end{document}